\documentclass[num-refs]{wiley-article}
\usepackage{siunitx}
\usepackage{comment}
\usepackage{subcaption}
\usepackage[export]{adjustbox}
\usepackage{lipsum}
\usepackage{amsmath}
\usepackage{mathtools}
\definecolor{ForestGreen}{rgb}{0.0, 0.5, 0}
\newcommand{\blue}{\color{black}}

\newcommand{\black}{\color{black}}

\papertype{Original Article}
\paperfield{Journal Section}

\title{Wake steering of multirotor wind turbines}

\author[1]{Gustav A. Speakman}
\author[2]{Mahdi Abkar}
\author[3]{Luis A. Mart{\'\i}nez-Tossas}
\author[1]{Majid Bastankhah}

\affil[1]{Department of Engineering, Durham University, Durham DH1 3LE, United Kingdom}
\affil[2]{Department of Mechanical and Production Engineering, Aarhus University, 8000 Aarhus C, Denmark}
\affil[3]{National Renewable Energy Laboratory, Golden, Colorado, USA}

\corraddress{Majid Bastankhah, Department of Engineering, Durham University, Durham DH1 3LE, United Kingdom}
\corremail{majid.bastankhah@durham.ac.uk}

\begin{document}

\maketitle

\begin{abstract}
\small
In this paper, wake steering is applied to multirotor turbines to determine whether it has the potential to reduce wind plant wake losses. Through application of rotor yaw to multirotor turbines, a new degree of freedom is introduced to wind farm control such that wakes can be expanded, channelled, or redirected to improve inflow conditions for downstream turbines. Five different yaw configurations are investigated (including a baseline case) by employing large-eddy simulations (LES) to generate a detailed representation of the velocity field downwind of a multirotor wind turbine. Two lower-fidelity models from single-rotor yaw studies (curled-wake model and analytical Gaussian wake model) are extended to the multirotor case and their results are compared with the LES data. For each model, the wake is analysed primarily by examining wake cross sections at different downwind distances. Further quantitative analysis is carried out through characterisations of wake centroids and widths over a range of streamwise locations, and through a brief analysis of power production. Most significantly, it is shown that rotor yaw can have a considerable impact on both the distribution and magnitude of the wake velocity deficit, leading to power gains for downstream turbines. 
 The lower-fidelity models show small deviation from the LES results for specific configurations; however, both are able to reasonably capture the wake trends over a large streamwise range.

% Please include a maximum of seven keywords
\end{abstract}

%\linenumbers

%CUT FROM ABSTRACT: Significant reductions in velocity deficit are observed for yaw arrangements which bring about high wake expansion or large overall wake deflections, leading to power gains for downstream turbines. 
%It is also shown that this technique has the capacity to channel turbine wakes to become narrower. 

\section{Introduction}\label{sec:intro}

As the wind energy sector continues to expand, it is of increasing relevance to optimise and improve turbine and plant technologies. Currently, in terms of overall power production, the most significant source of loss in a wind plant is the impact of upstream rotor wakes on downstream machines \citep{stevens2017flow,meneveau2019big}. Wakes of turbine rotors are typically characterised by velocity deficit and high levels of turbulence, as detailed by prior studies \citep[see the review of][and references therein]{porte2020wind}. Consequently, turbines operating in these wakes are not able to extract as much energy as their upstream counterparts, which tends to result in a reduced overall plant power output \citep{vermeer2003wind}.

To mitigate power losses in wind farms, it has been proposed that wake steering may offer an appropriate solution. This technique involves yawing upstream rotors such that wake flows are directed away from downstream turbines. While this leads to a decrease in upstream turbine power, this is outweighed by an increase in power output downstream caused by lower velocity deficit \citep{fleming2015simulation}. The potential of this technique has been examined in a number of studies such as \citet{jimenez2010application}, in which the effectiveness of wake steering  is demonstrated using large-eddy simulations (LES), and a simple analytical model is developed to describe wake deflection. \citet{gebraad2016wind} conducted an alternative analysis by adapting the ``top-hat'' profile proposed by \citet{jensen1983note} and optimising rotor yaw angles within a wind plant using game theoretic methods. More recently, \citet{bastankhah2016experimental} built on earlier findings \cite{bastankhah2014new}, employing a Gaussian distribution to describe velocity deficit. Details of cross-stream components of wake velocity have been examined by \citet{martinez2019aerodynamics}, in which a ``curled-wake'' model is formulated, and by \citet{shapiro2018modelling} in which yawed turbines are viewed as lifting lines. In recent developments, several wind tunnel studies \citep[e.g.,][]{bastankhah2019wind, macri2020experimental,campagnolo2020wind,hulsman2020turbine}, numerical simulations \citep[e.g.,][]{archer2019wake,King2020, gomez2020numerical,wang2020does}, as well as field studies \citep[e.g.,][]{fleming2017field, fleming2019initial, fleming2020continued, howland2019wind,doekemeijer2020field,brugger2020lidar} have examined wake steering for wind farms of single-rotor wind turbines, in which improvements have been seen in terms of both power output and reliability.

Another obstacle facing the industry from a more structural and logistical standpoint is that associated with the ``square-cube law,'' expounded primarily by  \citet{jamieson2012multi}. This problem essentially stems from the fact that energy extracted from a rotor scales with its area (proportional to square of diameter), but the production cost scales approximately with volume (proportional to cube of diameter). With the objective of lowering cost of energy, rotor sizes have therefore rapidly increased in recent years, though marginal gains have become steadily smaller due to higher manufacturing costs \citep{noyes2020analytic}. Further problems associated with large rotors, such as high shipping, assembly, and maintenance costs also indicate that the practice of rotor upscaling may be reaching its limit \citep{jamieson2014considerations}.

In tackling the problems associated with rotor upscaling, a promising solution is that of multirotor turbines. This concept involves installing a number of smaller rotors on a single support structure, rather than just one large rotor \citep{chasapogiannis2014analysis, ferede2020modal, vali2020rejecting, kirchner2020multi}. Primarily, this alleviates the growing expense of rotor manufacture since the total material volume associated with the smaller rotor components is less, leading to lower production costs. \citet{jamieson2012multi} also showed that further cost reductions are associated with installation and shipping, leading to overall savings estimated at 30\% for a \SI{20}{\mega\watt} turbine. Advantages are also seen from a reliability standpoint, since multirotor turbines are still able to operate at a reduced capacity when one generator fails, as noted by \citet{jamieson2014structural}. Where failure of a single-rotor turbine generator would constitute a significant loss to the overall power output of a wind farm, failure of one multirotor generator would be less severe since it only makes up a fraction of the turbine's overall capacity.

Further to this, one of the key merits of multirotor designs is a lower susceptibility to wind veer, an effect caused by the presence of the Coriolis force in the atmospheric boundary layer which results in varying wind direction with height \citep{Abkar2014, allaerts2015large, Lundquist2015, Abkar2016b, Xie2017}. For current large turbine designs the variation in direction across the rotor is nontrivial, and efficiency losses typically result since the wind direction is not normal to the rotor at all points within the swept area \citep{bhaganagar2015effects}. Wind veer also leads to the generation of a skewed wake which can affect the efficiency of downwind turbines \citep{abkar2018analytical, van2017coriolis}, and can even impact on interactions between wind farms \cite{van2015predicting}. In multirotor turbines, however, these effects are less pronounced, primarily due to smaller rotor sizes and the ability to yaw individual rotors such that they are properly adjusted to the wind direction at a given height. 

Finally, multirotor turbines are able to deliver an improved power output when compared with single-rotor equivalents. Field measurements and numerical simulations carried out by \citet{van2019power} showed a 1.8 $\pm$ 0.2\% increase in power production of the Vestas 4R-V29 demonstrator, an improvement attributed primarily to rotor interaction. When tested in wind farm arrays, \citet{van2019improved} estimated an increase of 0.3-1.7\% in annual energy production (AEP) for a 4-by-4 (16-turbine) arrangement. These AEP gains were primarily found for layout-aligned wind directions and for the first downstream wind turbine in a row. Such power increases are primarily seen as a result of faster wake recovery in the range of 5-8 rotor diameters, which is the typical streamwise turbine spacing for most wind farms. \citet{bastankhah2019multirotor} found that this was due to individual rotor wakes remaining distinct in this range, where at greater distances they merge to form a single wake. LES studies of \citet{ghaisas2018large} also demonstrated lower wake losses, which was linked to a higher planform energy flux and greater flow entrainment in the wake. Further to this, \citet{ghaisas2020effect} showed that wake losses were further reduced with increases in rotor tip spacing. Such improvements in wake recovery lead to lower velocity deficits downstream, facilitating greater power production from downwind machines.

\begin{figure*}[t]
\centering
  \begin{subfigure}[b]{0.46\textwidth}
    \includegraphics[width=\textwidth]{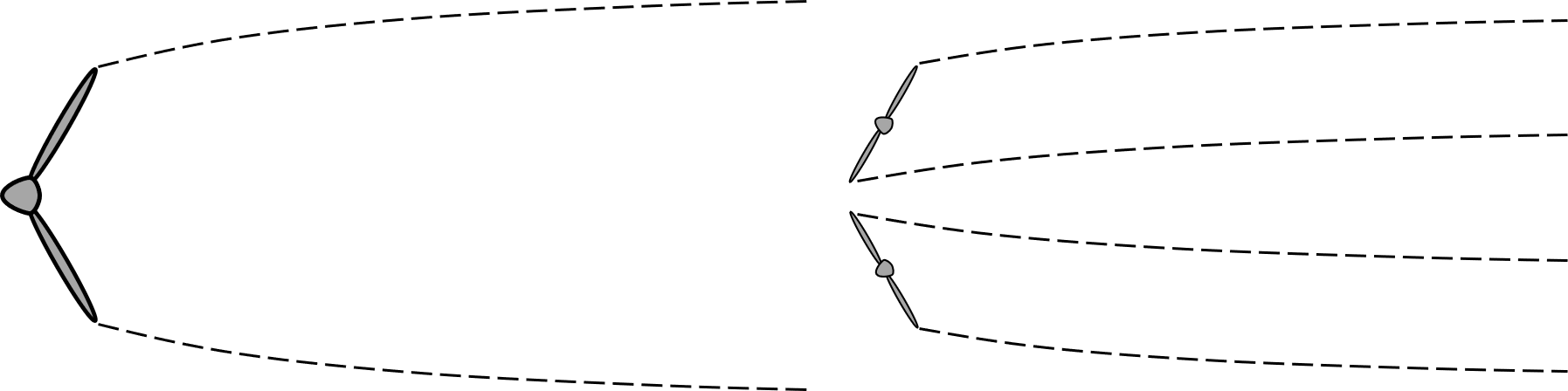}
    \caption{Wake expansion by coning (left) and multirotor yaw (right)}
    \label{fig:div_cone_yaw}
  \end{subfigure}
  \hspace*{5mm}
  ~
  \begin{subfigure}[b]{0.46\textwidth}
    \includegraphics[width=\textwidth]{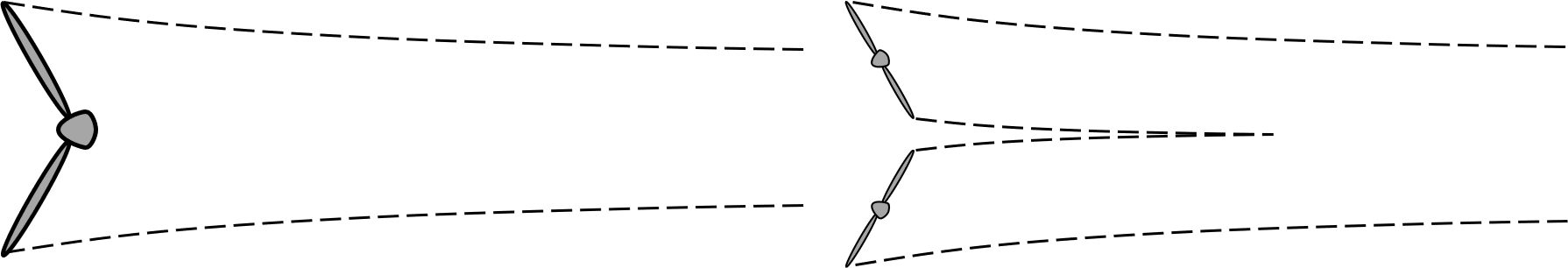}
    \vspace*{-2mm}
    \caption{Wake channelling by coning (left) and multirotor yaw (right)}
    \label{fig:con_cone_yaw}
  \end{subfigure}
\caption{Introduction of wake control through coning and multirotor yaw. Fluid velocity is in the left to right direction. The comparison of these technologies is made due to the similarities in their effects, but coning is not examined in this paper.}
    \label{fig:cone_yaw_sch}
\end{figure*}

This paper aims to bring together advances in both wake steering and multirotor research to introduce a new degree of freedom to wind farm control. It is proposed that wake steering is applied to individual rotors of a multirotor turbine such that the wake can be expanded, channelled, or redirected to reduce downstream losses. In regards to wake modulation, a yawed multirotor turbine can be compared with rotor coning, in which single-rotor turbine blades are angled in the streamwise direction, as illustrated in Fig. \ref{fig:cone_yaw_sch}. While coning is primarily a load alignment strategy in which cantilever blade loads are converted to tensile ones \citep{bortolotti2019comparison}, it can be seen in Fig. \ref{fig:cone_yaw_sch} that this method also allows some control of wake expansion. However, since this design is still very much at a conceptual stage, there exist several practical issues surrounding its implementation. Details of rotor mounting may be challenging, and the extent to which benefits of a fully redesigned turbine outweigh its cost have been questioned \citep{crawford2007path}. By comparison, rotor yaw is already a mature technology, currently utilised in adjusting single-rotor turbines to the incoming wind. Therefore, rotor coning is not examined in this paper; however, what is proposed is that multirotor turbines may be able to deliver a similar type of wake control.

The method suggested to achieve this is individual rotor yaw since it is a well established technique, however, the effects of rotor tilt are also of interest \cite{cossu2020evaluation,scott2020characterizing} and could be examined in further works. \black The effects of pitch and roll in floating turbines may also be applied to the multirotor case, as has been investigated in single rotor wind tunnel studies \citep[e.g.,][]{fu2019wake,fu2020phase,rockel2014experimental,rockel2017dynamic} and numerical studies of \citet{xiao2019large}. More than this, the impacts of different atmospheric boundary layers \cite{dorenkamper2015impact,dorenkamper2015offshore,witha2014large} and wave structures \cite{yang2014large} on turbine wakes may also be extended to multirotor studies.
\black

A number of different multirotor yaw configurations are modelled in this paper, evaluated qualitatively by examining wake cross sections and quantitatively by characterising wake centroid locations and wake widths. A final assessment of each yaw configuration is achieved through an analysis of power production. LES data forms the basis of the results, which is compared with a Gaussian analytical wake model and a curled-wake model. The formulation and application of each of these models is explained in \S2. The results are presented in \S3, with an exact description of the model setup and tested configurations. Wake cross sections are first presented and examined to investigate the nature of wake development in each case. Variation of rotor-averaged quantities with downwind distance is also presented alongside this data. In \S \ref{sec:yc_Sy}, the wake centroids and widths are mathematically characterised and compared for all models to further quantify the merits of each yaw configuration. Finally, \S \ref{sec:power} presents a power analysis as a closing assessment of this multirotor yaw scheme.

\section{Methodology} \label{sec:meth}

\subsection{LES Model} \label{sec:les}
\begin{figure}[t]
\centering
  \begin{subfigure}[b]{0.4\textwidth}
    \includegraphics[width=\textwidth]{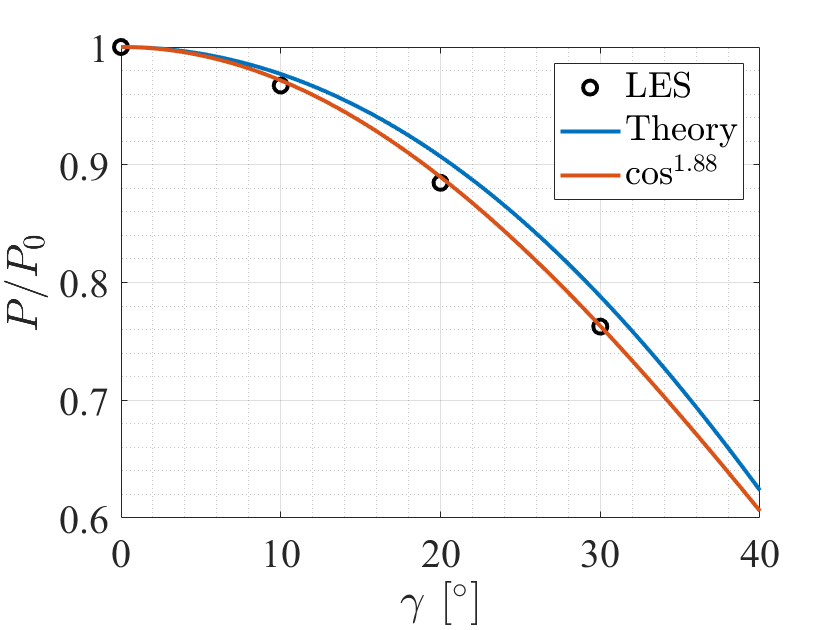}
    \caption{Variation of normalised power, $P/P_0$, with yaw angle, $\gamma$.}
    \label{fig:PP0}
  \end{subfigure}
  \hspace*{1em}
  ~
  \begin{subfigure}[b]{0.4\textwidth}
  \includegraphics[width=\textwidth]{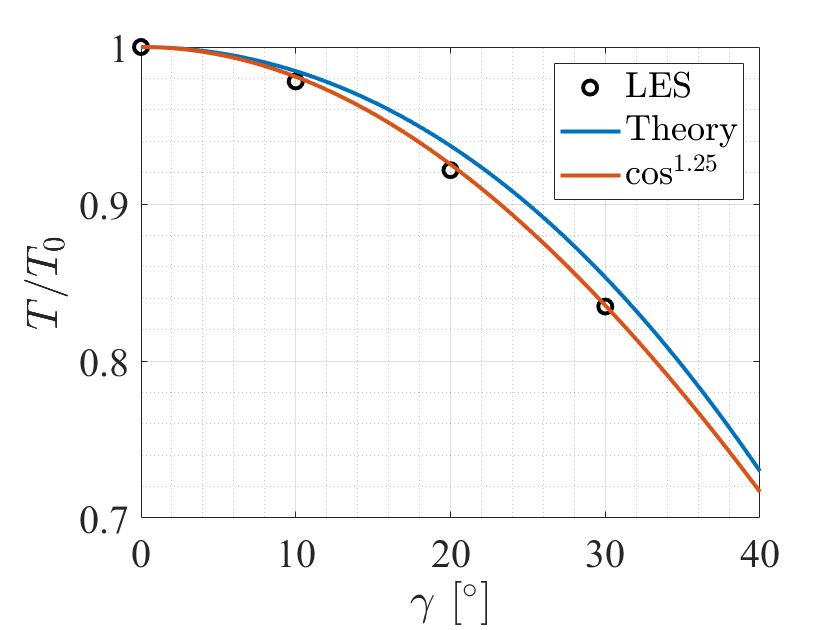}
    \caption{Variation of normalised thrust, $T/T_0$, with yaw angle, $\gamma$}
    \label{fig:TT0}
  \end{subfigure}
  \caption{Variation of normalised power, $P/P_0$, and normalised thrust, $T/T_0$, with yaw angle, $\gamma$. The blue line is calculated according to Equations. 4 and 5 and black circles represent LES data. The red line is a power function fitted to LES data.}
    \label{fig:PP0_TT0}
\end{figure}

The LES code described here solves the filtered continuity and momentum equations for an incompressible turbulent fluid as 

\begin{equation}\label{eq:NS}
\frac{\partial {u}_i}{\partial x_i}=0,~~~ \frac{\partial{u}_i}{\partial t}+
{u}_j \frac{\partial {u}_i}{\partial x_j} =
-\frac{1}{\rho_o} \frac{\partial {p}}{\partial x_i}
-\frac{\partial \tau_{ij}}{\partial x_j} 
-\frac{f_i}{\rho_o},
\end{equation} 
where $({u}_1,{u}_2,{u}_3)=({u},{v},{w})$ 
are the filtered/resolved velocity components, where $i=1,2$ and $3$ indicate the streamwise $x$, spanwise $y$, and vertical $z$ directions, respectively.
The filtered/resolved pressure is denoted by ${p}$.  
$t$ is time, $\rho_o$ is the fluid density, $\tau_{ij}$ is the subfilter stress tensor, and $f_i$ represents wind turbines' effects on the air flow. 
The code utilizes a second-order finite-difference discretization in the vertical direction together with a pseudo-spectral method in the horizontal directions. The time integration is carried out using a second-order Adam-Bashforth method. The molecular viscous forces are neglected in the momentum equation; hence the flow is at a nominally infinite Reynolds number. 
In the code, the subfilter stress tensor is parameterized using  the Lagrangian scale-dependent dynamic model \citep{Stoll2006a}. The instantaneous wall shear stress is computed based on the local application of Monin–Obukhov similarity theory \citep{Bou-Zeid2005,Yang2018}. %Moeng1984,Bou-Zeid2005,
In order to generate the inflow condition for the wake flow simulations, a precursor technique is used in which a fully developed boundary-layer flow under neutral condition is simulated. % and is fed into the wake simulations. 
The size of the computational domain is \SI{1600}{\meter} $\times$ \SI{800}{\meter} $\times$ \SI{355}{\meter}, and it is  discretized uniformly into $160\times 160\times 72$ computational grid points in the $x$, $y$, and $z$ directions, respectively. The boundary-layer flow is driven by an imposed pressure gradient. 
The effective surface roughness height is set to \SI{0.005}{\meter}.
%In the wake flow simulation, a fringe zone is used to adjust the flow from the downstream wake state to an undisturbed inflow condition. % \citep{Abkar2016,Munters2015}. %Munters2015 
The turbine-induced forces are modelled using the standard non-rotational actuator-disk method be described by \citet{Calaf2010} using the wind velocity at the rotor plane and the disk-based thrust coefficient, $C_T'$. Values of $C_T'$ and the nominal turbine thrust coefficient, $C_T$, are related to the turbine thrust force, $T$, as follows

\begin{equation}\label{eq:thrust}
    T=\frac{1}{2}\rho_0 A C_T' \bar{u}_d^2=\frac{1}{2}\rho_0 A C_T \bar{u}_{\infty}^2 \cos^2{\gamma},
\end{equation}
where $A$ is the rotor area, and $\bar{u}_d$ is the time-averaged normal velocity at the rotor. The time-averaged upstream undisturbed velocity is denoted by $\bar{u}_{\infty}$, and $\gamma$ is the yaw angle. The thrust force is distributed uniformly over the rotor area, and the same value of $C_T'=4/3$ is used for all simulations. Similarly, for power 

\begin{equation}\label{eq:power}
    P=\frac{1}{2}\rho_0 A C_P' \bar{u}_d^3=\frac{1}{2}\rho_0 A C_P \bar{u}_{\infty}^3 \cos^3{\gamma},
\end{equation}
where $C_P'$ is the disk-based power coefficient and $C_P$ is the nominal turbine power coefficient. Note that $C_P'$ and $C_T'$ are equivalent, that is $C_P' = C_T' = 4/3$. From the theoretical work of \citet{shapiro2018modelling}, the relationship between $C_T$ and $C_T'$ is defined as
\begin{equation} \label{eq:CT}
    C_T= C_T'\left(\frac{4}{4+C_T'\cos^2{\gamma}}\right)^2.
\end{equation}
We use a similar approach here to show that $C_P$ and $C_P'$ are related to each other as follows
\begin{equation} \label{eq:CP}
    C_P= C_P'\left(\frac{4}{4+C_P'\cos^2{\gamma}}\right)^3.
\end{equation}
Using the above theoretical relationships, the normalised power, $P/P_0$, and normalised thrust, $T/T_0$, are plotted in Fig. \ref{fig:PP0_TT0}, where $P_0\!=\!P(\gamma\!=\!0)$ and $T_0\!=\!T(\gamma\!=\!0)$. The figure shows how much power is lost from a turbine as a result of yawing rotors. LES values and a fitted power function are also plotted on the same figure. For the power function ($\cos^n$), values of $n$ were found to be 1.88 for $P/P_0$ and 1.25 for $T/T_0$. LES values are taken from a single-rotor simulation, using the same framework detailed above and are used throughout this paper as inputs to the Gaussian model and for power calculations in \S \ref{sec:power}. The LES framework described here has been well validated and used in earlier wind energy research publications. The reader may refer to Refs. \citep{Wu2011,Abkar2014,Abkar2016b} for a more detailed description of the LES framework and the solver.

%The disk-based thrust coefficient of the turbines $C_t'$ is set to $4/3$. %$C_t'$ 
%Similar to the earlier studies (e.g. \citet{allaerts2015large,Abkar2015}), 
%the disk-based thrust coefficient of the turbines ($C_t$) is set to $0.75$.  
%The total thrust force acting on the turbine is modeled as 
%\begin{equation}\label{eq:Thrust}
%T=\frac{1}{2}\rho_o C_t'A_d{\tilde{M}_d}^2,  
%\end{equation}
%where $A_d$ is the disk area, and $\tilde{M}_d$ is the velocity aligned with the turbine axes at the rotor plane. 

%In the wake flow simulation, a fringe zone is used to adjust the flow from the downstream wake state to an undisturbed inflow condition \citep{Abkar2016,Munters2015}. %Munters2015 
%The vertical profiles of mean wind speed and turbulence intensity in the  streamwise  direction obtained from the precursor simulation are shown in Fig. \ref{Fig_BL}. 

\black \subsection{Gaussian Analytical Model} \label{sec:gauss}

\begin{figure}[t]
\centering
  \begin{subfigure}[t]{0.2\textwidth}
    \includegraphics[width=\textwidth]{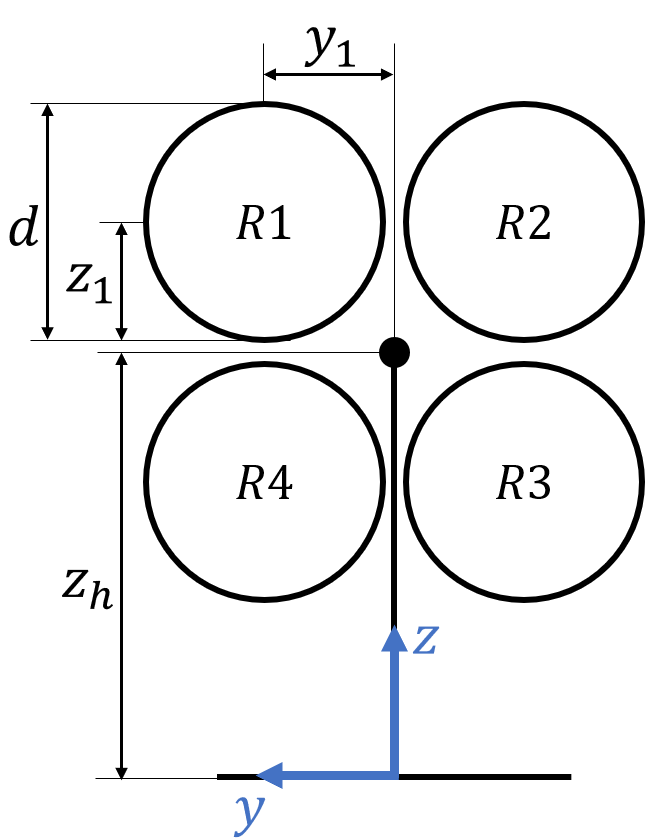}
    \caption{multirotor turbine in the y-z plane from an upstream perspective.}
    \label{fig:name_conv}
  \end{subfigure}
  \hspace*{1em}
  ~
  \begin{subfigure}[t]{0.24\textwidth}
  \includegraphics[width=\textwidth]{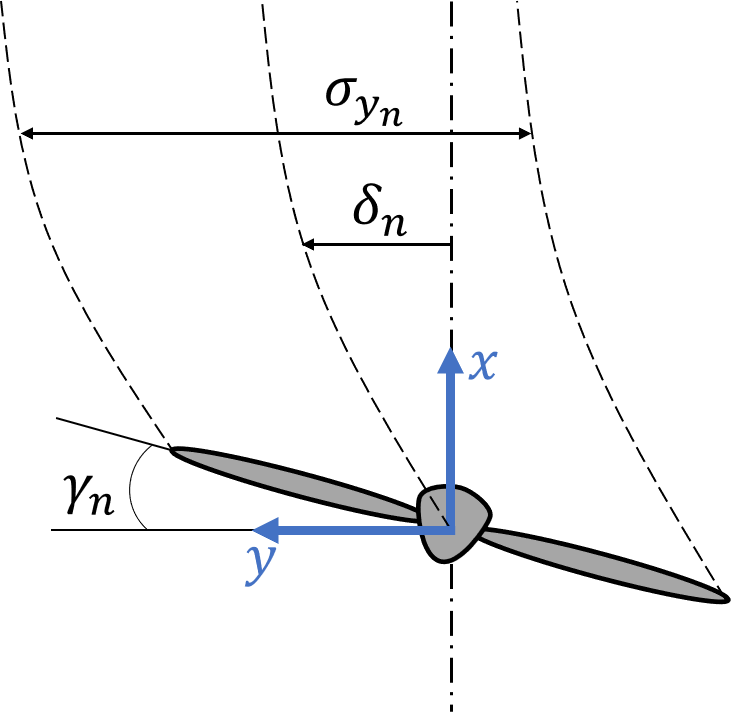}
      \caption{Positive yaw of a rotor in the clockwise direction, illustrated in the x-y plane from above.}
    \label{fig:yaw_sch}
  \end{subfigure}
  \caption{Schematics of a yawed multirotor turbine.}
    \label{fig:MR_sch}
\end{figure}

The analytical model employed is an extension of the Gaussian wake model developed by \citet{bastankhah2016experimental}, which was initially developed for single yawed rotors. In the case of a multirotor turbine, such as that illustrated in Fig. \ref{fig:name_conv}, the Gaussian wake model is applied to each rotor, and the individual wakes are then linearly superposed as suggested by \citet{bastankhah2019multirotor}. \black For the $n^{th}$ yawed rotor, the individual wake widths in the lateral and vertical directions respectively, can be found by

\begin{equation} \label{eq:sr_width}
\large
\begin{cases}
    \frac{\sigma_{y_n}}{d} = k_n\frac{x - x_0}{d} + \frac{\cos{\gamma_n}}{\sqrt{8}}, \\
    \frac{\sigma_{z_n}}{d} = k_n\frac{x - x_0}{d} + \frac{1}{\sqrt{8}},
    \end{cases}
\end{equation}
where $k_n$ is the wake growth rate associated with each rotor, which is assumed to be equal in lateral and vertical directions. Lateral and vertical dimensions are denoted by $y$ and $z$, respectively, and widths are normalised by rotor diameter, $d$. Streamwise distance from the turbine is denoted by $x$, and $x_0$ is the streamwise distance at which the onset of the far wake region occurs, as detailed by \citet{bastankhah2016experimental}. The rotor yaw angle is $\gamma_n$, as displayed in Fig. \ref{fig:yaw_sch}, defined as positive in the anticlockwise direction when viewed from above. The maximum velocity deficit associated with each rotor, $C_n$, may be given by

\begin{equation} \label{eq:max_vel}
C_n = 1 - \sqrt{1 - \frac{C_T\cos^3(\gamma_n)}{8\sigma_{y_n}\sigma_{z_n}/d^2}},
\end{equation} 
where $C_T$ is the rotor thrust coefficient. Note that the reason for the apparent discrepancy between Equation. \ref{eq:max_vel} and the one in the original work is a different definition of $C_T$ adopted in the current study (Equation. \ref{eq:CT}). In fact, the value of $C_T\cos^2{\gamma}$ in the current study is equivalent to $C_T$, as defined in \citet{bastankhah2016experimental}. It is assumed that thrust coefficient and rotor diameter are constant for all rotors. The maximum value, $C_n$, may be used to describe the velocity deficit distribution as a three-dimensional Gaussian profile, given by

\begin{equation} \label{eq:sr_veldef}
\large
 \frac{\Delta \bar{u}_n(x,y,z)}{\bar{u}_h} = C_n e^{\frac{(y-y_n-\delta_n)^2}{2\sigma_{y_n}^2}}  e^{\frac{(z-z_h-z_n)^2}{2\sigma_{z_n}^2}},
\end{equation}
where $\Delta \bar{u}_n(x,y,z)$ is the time-averaged velocity deficit of a single-rotor, normalised by the time-averaged inflow velocity at hub height, $\bar{u}_h$. The lateral and vertical rotor offsets from the turbine centre are denoted by $y_n$ and $z_n$ respectively, where $z_h$ is the turbine hub height and $\delta_n$ is the lateral wake deflection due to yaw. \black The total normalised velocity deficit distribution of the multirotor turbine is given by the linear sum of the contributions from each rotor, hence

\begin{equation} \label{eq:mr_veldef}
\frac{\Delta \bar{u}(x,y,z)}{\bar{u}_h} = \sum_{i=1}^{n} \frac{\Delta \bar{u}_n(x,y,z)}{\bar{u}_h}.
\end{equation}
The above sum of individual rotor velocity deficit contributions facilitates generation of a full flow field which can be interrogated to examine velocity characteristics and features of the wake expansion. 

Finally, the spanwise velocity distribution, $\bar{v}(x,y,z)$, may be found by calculating the product of the streamwise velocity field, $\bar{u}(x,y,z)$, and the skew angle distribution, $\theta(x,y,z)$. Streamwise velocity may be found directly as $\bar{u}(x,y,z) = \bar{u}_h\left(1-\Delta \bar{u}(x,y,z)\right)$, and skew angle distribution for the $n^{th}$ rotor can be computed from the Gaussian profile suggested by \citet{bastankhah2016experimental},

\begin{equation} \label{eq:theta}
\large
\frac{\theta_n(x,y,z)}{\theta_m} = e^{\frac{-(y-y_n-\delta_n+\sigma_{y_n})^2}{2 \sigma_{y_n}^2}} e^{\frac{-(z-z_h-z_n)^2}{2 \sigma_{z_n}^2}},
\normalsize
\end{equation}
where $\theta_m$ is the maximum skew angle at each downwind location. The interested reader is referred to the original work for more information. The total spanwise velocity caused by all four rotors is obtained by linear superposition of each rotor contribution, akin to the one for the velocity deficit. \black

\subsection{Curled-Wake Model} \label{sec:CW}
The curled-wake model uses a simplified
version of the 
Reynolds-averaged
Navier Stokes (RANS) equations for 
the velocity deficit of
wind turbines in yaw \citep{martinez2019aerodynamics}.
The streamwise component of the simplified 
RANS equation is

\begin{equation} \label{eq:curled_wake}
    (U + \Delta \bar{u}) \, 
    \frac{\partial \Delta \bar{u}}{\partial x} +
    V
    \frac{\partial \Delta \bar{u}}{\partial y} +
    W
    \frac{\partial \Delta \bar{u}}{\partial z}
    = 
     \nu_{\rm eff}
     \left(
      \frac{\partial^2 \Delta \bar{u}}{\partial y^2}
     + \frac{\partial^2 \Delta \bar{u}}{\partial z^2}
     \right),
\end{equation}
where
$V$ and $W$ are the spanwise velocities from the analytical formulations caused by yaw,
$U$ is the inflow streamwise velocity, and 
$\nu_{\rm eff}$ is the turbulent viscosity.
Equation. \ref{eq:curled_wake} is a parabolic equation which
is solved numerically using a ``forward-time centered-space'' method \citep{martinez2019aerodynamics}.
The initial condition for the wake deficit at the location of a turbine is computed from axial momentum theory based on the thrust coefficient (same as the one in Equation. \ref{eq:sr_veldef} used in the Gaussian model when $x=0$).
For the case of a multirotor, each turbine wake is 
initialized and the superposition of the wakes
is done explicitly by solving Equation. \ref{eq:curled_wake}.
The numerical solution of Equation. \ref{eq:curled_wake}
provides the wake deficit for all the turbines
without the need to use a superposition method.
The eddy viscosity is modeled by scaling the mixing length from the atmospheric boundary layer to take into account the wake added turbulence \cite{Blackadar1962, martinez2019aerodynamics} as

\begin{equation} \label{eq:nu}
	\ell_{\rm m} = \frac{\kappa z }{\left(1+\kappa z / \lambda \right)},
	\qquad \qquad \qquad
	\nu_{\rm eff} = C \, \ell_{\rm m}^2 \left| \frac{d U}{dz}\right|.
\end{equation}
Equation \ref{eq:nu} shows the eddy-viscosity model where $\ell_{\rm m}$ is the mixing length, $z$ is the distance from the ground, $\lambda$ is the value of the mixing length in the free atmosphere, $\kappa$ is the von-K\'arm\'an constant, and $C=8$ is a tuned constant to take into account wake added turbulence \cite{Bay2020FlorisCurl}.

\section{Results}

\begin{figure*}[t]
\centering
  \begin{subfigure}[t]{0.18\textwidth}
    \includegraphics[width=\textwidth]{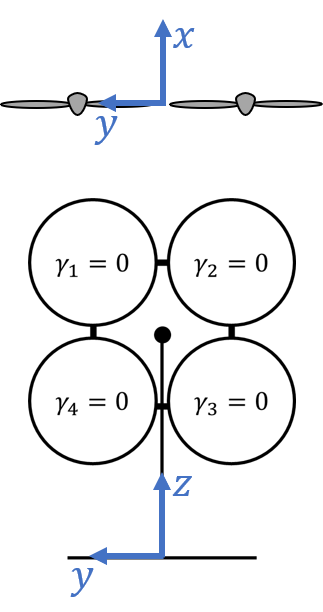}
    \caption{Zero yaw}
    \label{fig:z_yaw_sch}
  \end{subfigure}
  \hspace*{1.5mm}
  \begin{subfigure}[t]{0.18\textwidth}
    \includegraphics[width=\textwidth]{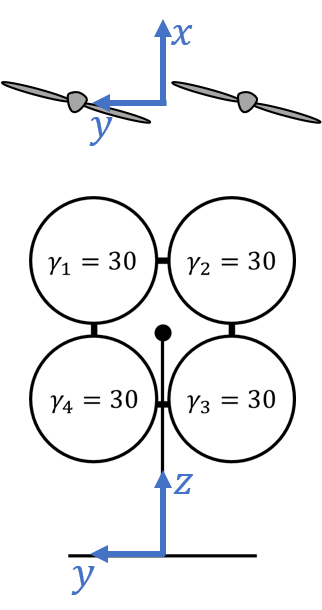}
    \caption{Equal yaw}
    \label{fig:eq_yaw_sch}
  \end{subfigure}
  \hspace*{1.5mm}
  \begin{subfigure}[t]{0.18\textwidth}
    \includegraphics[width=\textwidth]{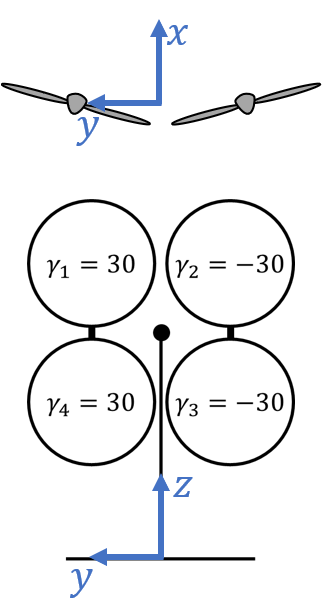}
    \caption{Divergent rotors}
    \label{fig:div_ver_sch}
  \end{subfigure}
  \hspace*{1.5mm}
  \begin{subfigure}[t]{0.18\textwidth}
    \includegraphics[width=\textwidth]{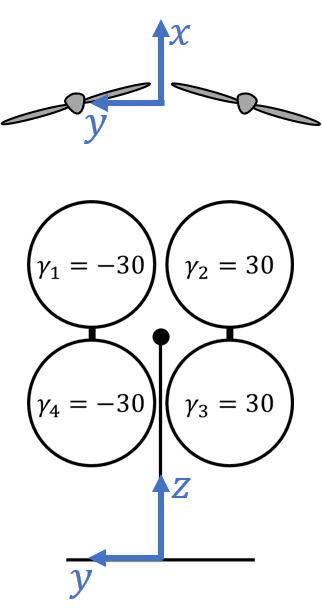}
    \caption{Convergent rotors}
    \label{fig:con_ver_sch}
  \end{subfigure}
  \hspace*{1.5mm}
  \begin{subfigure}[t]{0.18\textwidth}
    \includegraphics[width=\textwidth]{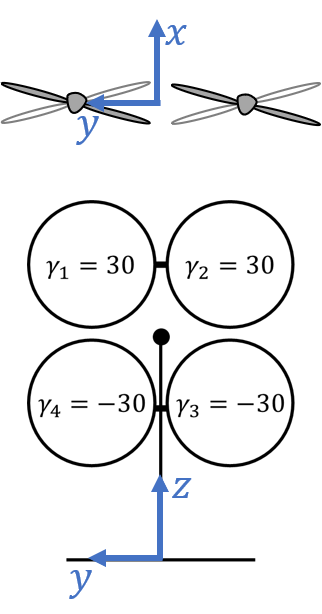}
    \caption{Crossed rotors}
    \label{fig:div_lat_sch}
  \end{subfigure}
  \caption{Schematics of the five tested yaw configurations viewed from above (top row) and from an upstream perspective (bottom row). Yaw angles are displayed on each rotor. Black solid lines connect rotors with the same yaw.}
    \label{fig:all_configs}
\end{figure*}

The model setup of the simulated turbine is illustrated in Fig. \ref{fig:MR_sch}, which shows four rotors labelled $R1$-$R4$ in a 2-by-2 arrangement. Each rotor is yawed by an angle, $\gamma_n$, resulting in a wake deflection, $\delta_n$, and an individual rotor wake width, $\sigma_{y_n}$.  The turbine hub height, $z_h$, is \SI{70}{\meter}, and the rotors have a diameter, $d$, of \SI{40}{\meter}. The swept area is equivalent to that of a single-rotor turbine with rotor diameter $D = $ \SI{80}{\meter}, since for an $n$-rotor turbine $d = D/\sqrt{n}$. Rotors have a tip spacing, $s$, of $0.1d$ (\SI{4}{\meter}) from one another. Rotor centres are therefore offset by \SI{22}{\meter} laterally and vertically from the turbine centre. At hub height, the mean incoming flow velocity, $\bar{u}_h$, is \SI{8}{\meter\per\second} and the streamwise turbulence intensity, $I_0$, is 6.7\%. Finally, the wake expansion rates for the Gaussian wake model were calculated using the relationship $k = 0.35I_0$ reported by \citet{carbajo2018wind}. Turbulence intensity values at the top and bottom rotor centres were used such that the top wake expansion rate was approximately 0.022, and the bottom wake expansion rate was 0.026. Note that for simplicity the wind veer is not modelled in this study, but as mentioned in \S\ref{sec:intro}, yawing multirotor turbines seems to be a promising method to maximise energy production in this case.

For the sake of brevity, the number of tested configurations was limited to the five cases illustrated in Fig. \ref{fig:all_configs}, where the yaw angles are indicated on each rotor. These include (a) a zero yaw case, (b) an equal yaw case, (c) divergent rotors, (d) convergent rotors, and (e) crossed rotors. The zero yaw case is the baseline, where no rotors are yawed, and the equal yaw case describes a configuration where all yaw angles are the same. ``Divergent rotors'' are so called because rotors appear divergent with respect to x, when viewed from above, as shown in Fig. \ref{fig:div_ver_sch}. Conversely, ``convergent rotors'' are so called because rotors appear convergent with respect to x, like that shown in Fig. \ref{fig:con_ver_sch}. In both of these cases, left rotors will have equal and opposite yaw angles to the right rotors. In other words, vertically adjacent rotors will have the same yaw. Finally, ``crossed rotors'' describes a case where rotors appear crossed from above. This is a result of top rotors having equal and opposite yaw angles to bottom rotors. Hence, laterally adjacent rotors will have the same yaw. Since the wake is relatively similar regardless of which way the rotors are crossed, only one crossed rotor case was studied. However, we note that there might be some differences between the two crossed rotor configurations in the case of wind veer. The four yawed cases were chosen based on their potential to deliver wake expansion, deflection, or channelling. It is out of the scope of this work to examine all possible configurations. However, we note that further work could be carried out to examine other promising configurations.

The magnitude of all rotor yaw angles was \SI{30}{\degree}, in order to clearly show yaw effects such as kidney-shaped (curled) wakes which have previously been observed by \citet{bastankhah2016experimental} and \citet{martinez2019aerodynamics}. Beyond \SI{30}{\degree}, wake steering has been found to have diminishing effects on wake deflection \citep{parkin2001application,jimenez2010application}, and negative effects on turbine loading \citep{kragh2014load}. All configurations involve yawing rotors in pairs, which are either vertically or horizontally adjacent. This choice was made so as to replicate the yaw mechanisms of utility-scale multirotor turbines such as the Vestas 4R-V29 \citep{van2019power}, which is able to apply a \SI{3}{\degree} toe-out angle to individual rotors. While this angle is small, it is a form of individual rotor yaw. Given that no experimental data are available and that the LES framework has previously shown good predictions of complex flow features, this model is taken as the closest approximation of real flow development. The LES has a high computational cost, however, which is why lower-fidelity models are used for comparison. Curled-wake predictions are obtained much more rapidly, though simplified RANS equations must still be solved. The Gaussian model is the least computationally expensive, only requiring the solution of analytical, closed-form equations. Results focus primarily on the far wake since the two lower-order models are not able to resolve the near-wake region.

\subsection{Multirotor Turbine Wake Characteristics} \label{sec:MR_turb_wake}

\begin{figure*}[t] %***ZERO YAW VELDEF***
  \includegraphics[width=\textwidth]{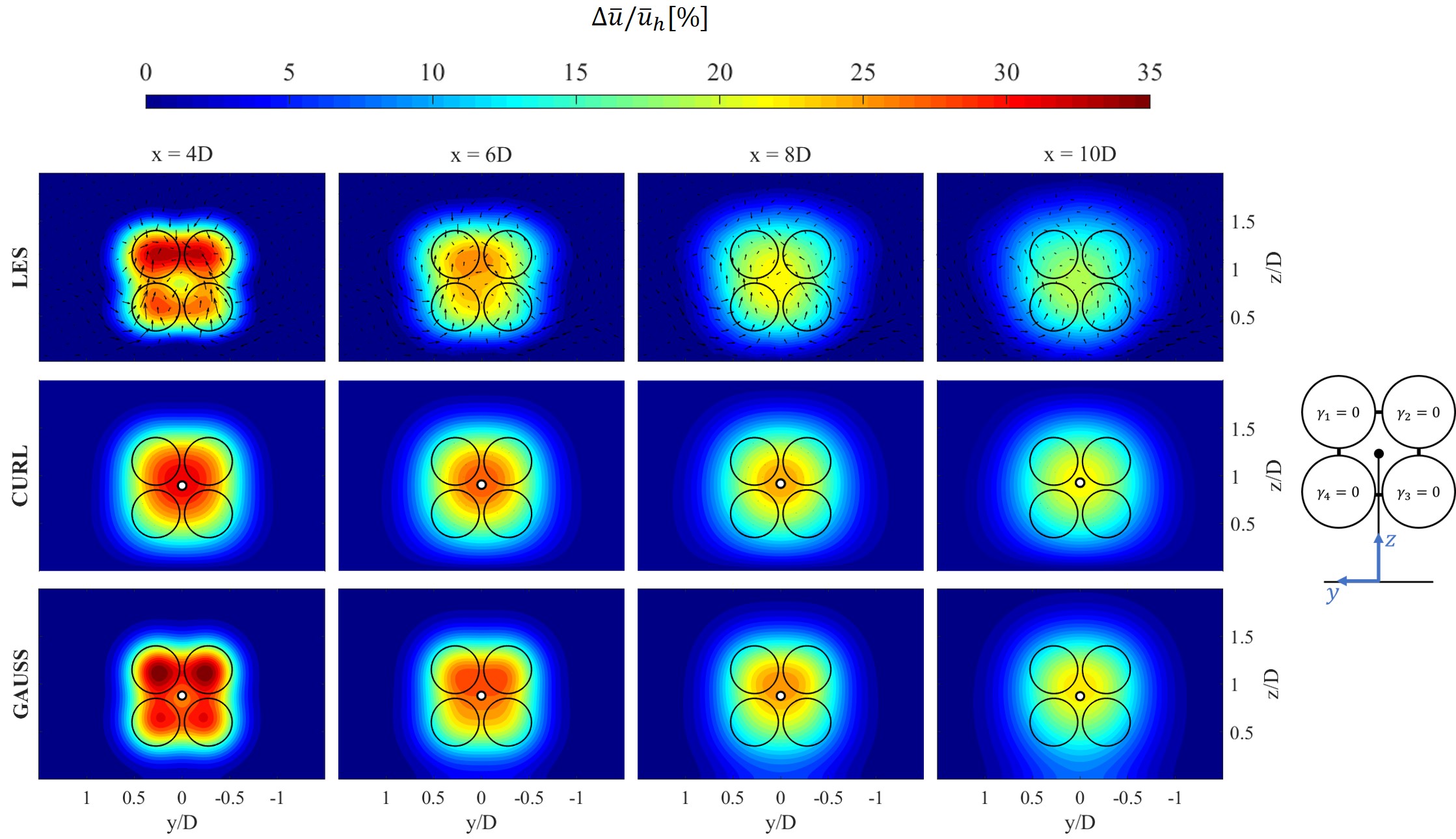}
  \caption{Zero yaw wake cross sections at four downwind distances. Contours show normalised velocity deficit, $\Delta \bar{u}/\bar{u}_h$, and vectors indicate cross-stream velocity. Black circles illustrate rotor swept area, and a white dot denotes the wake centroid.}
\label{fig:z_yaw_veldef}
%\vspace*{-5mm}
\end{figure*}

\begin{figure*}[t] %***EQUAL YAW VELDEF***
  \includegraphics[width=\textwidth]{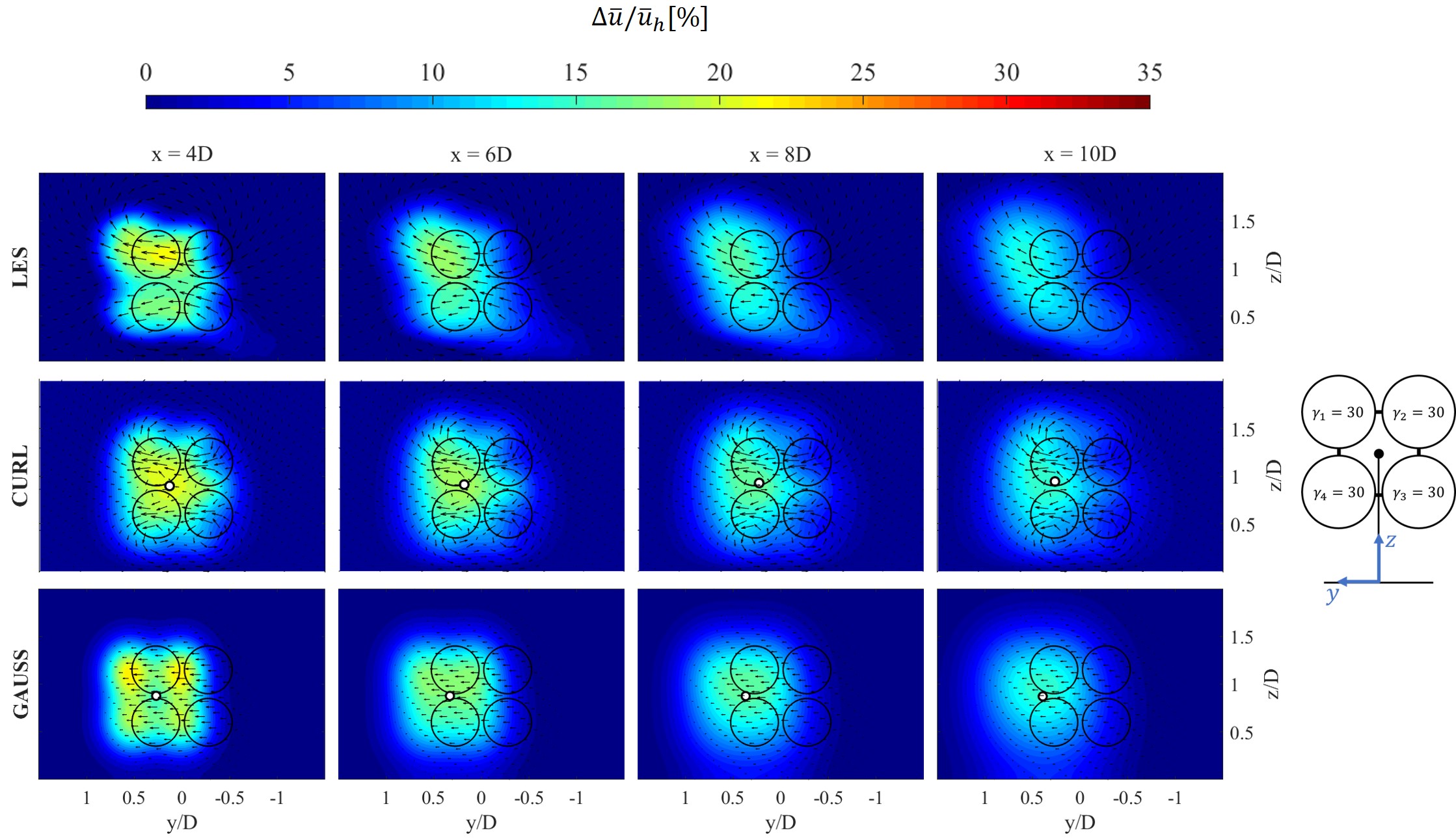}
  \caption{Equal yaw wake cross sections at four downwind distances. Contours show normalised velocity deficit, $\Delta \bar{u}/\bar{u}_h$, and vectors indicate cross-stream velocity. Black circles illustrate rotor swept area, and a white dot denotes the wake centroid.}
\label{fig:eq_yaw_veldef}
%\vspace*{-5mm}
\end{figure*}

\begin{figure*}[t] %***DIV_VER_VELDEF***
  \includegraphics[width=\textwidth]{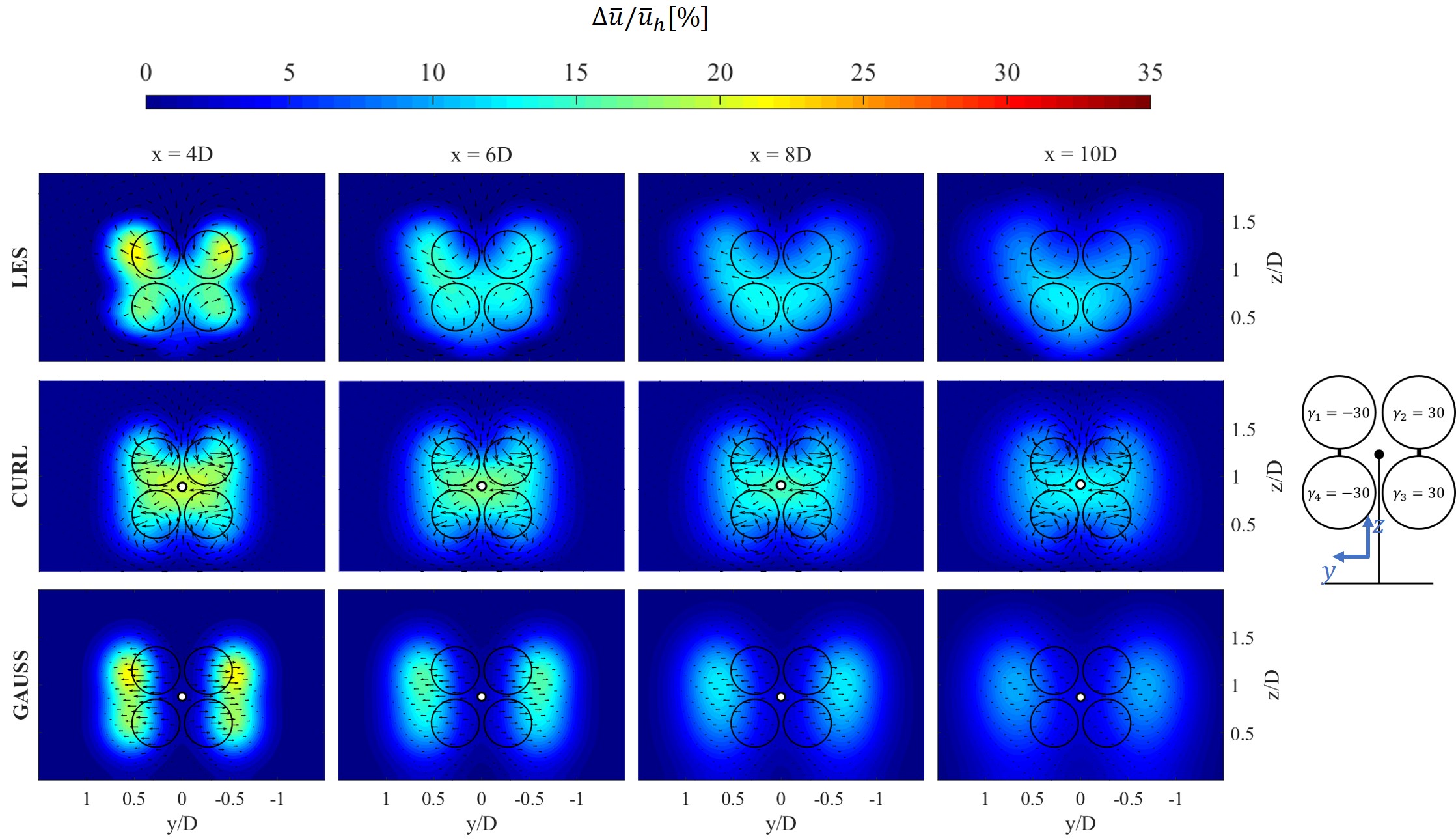}
  \caption{Divergent rotor wake cross sections at four downwind distances. Contours show normalised velocity deficit, $\Delta \bar{u}/\bar{u}_h$, and vectors indicate cross-stream velocity. Black circles illustrate rotor swept area, and a white dot denotes the wake centroid.}
\label{fig:div_ver_veldef}
%\vspace*{-5mm}
\end{figure*}

\begin{figure*}[t] %***CON_VER_VELDEF***
  \includegraphics[width=\textwidth]{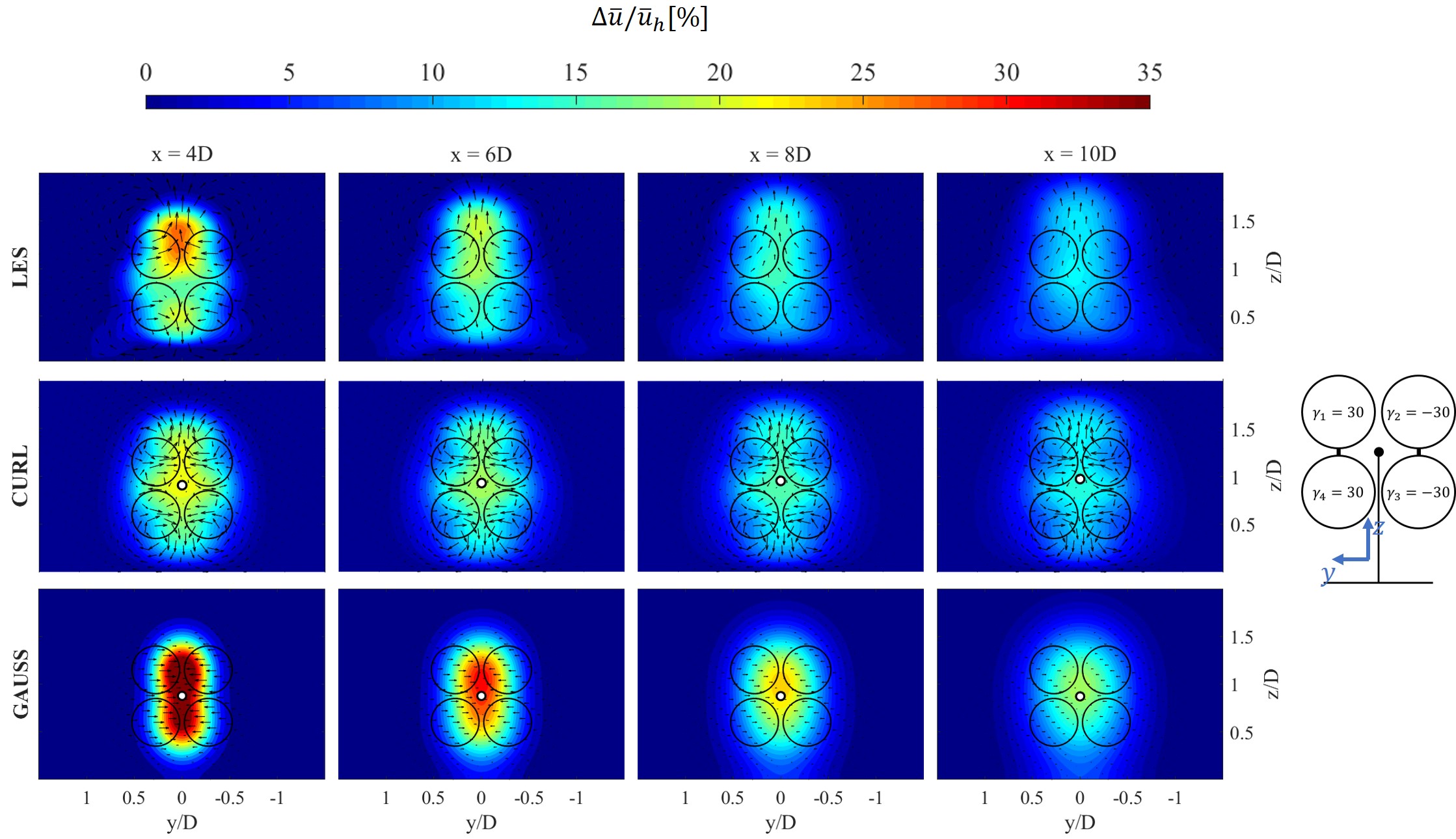}
  \caption{Convergent rotor wake cross sections at four downwind distances.  Contours show normalised velocity deficit, $\Delta \bar{u}/\bar{u}_h$, and vectors indicate cross-stream velocity. Black circles illustrate rotor swept area, and a white dot denotes the wake centroid.}
\label{fig:con_ver_veldef}
%\vspace*{-5mm}
\end{figure*}

\begin{figure*}[t] %***DIV_LAT_VELDEF***
  \includegraphics[width=\textwidth]{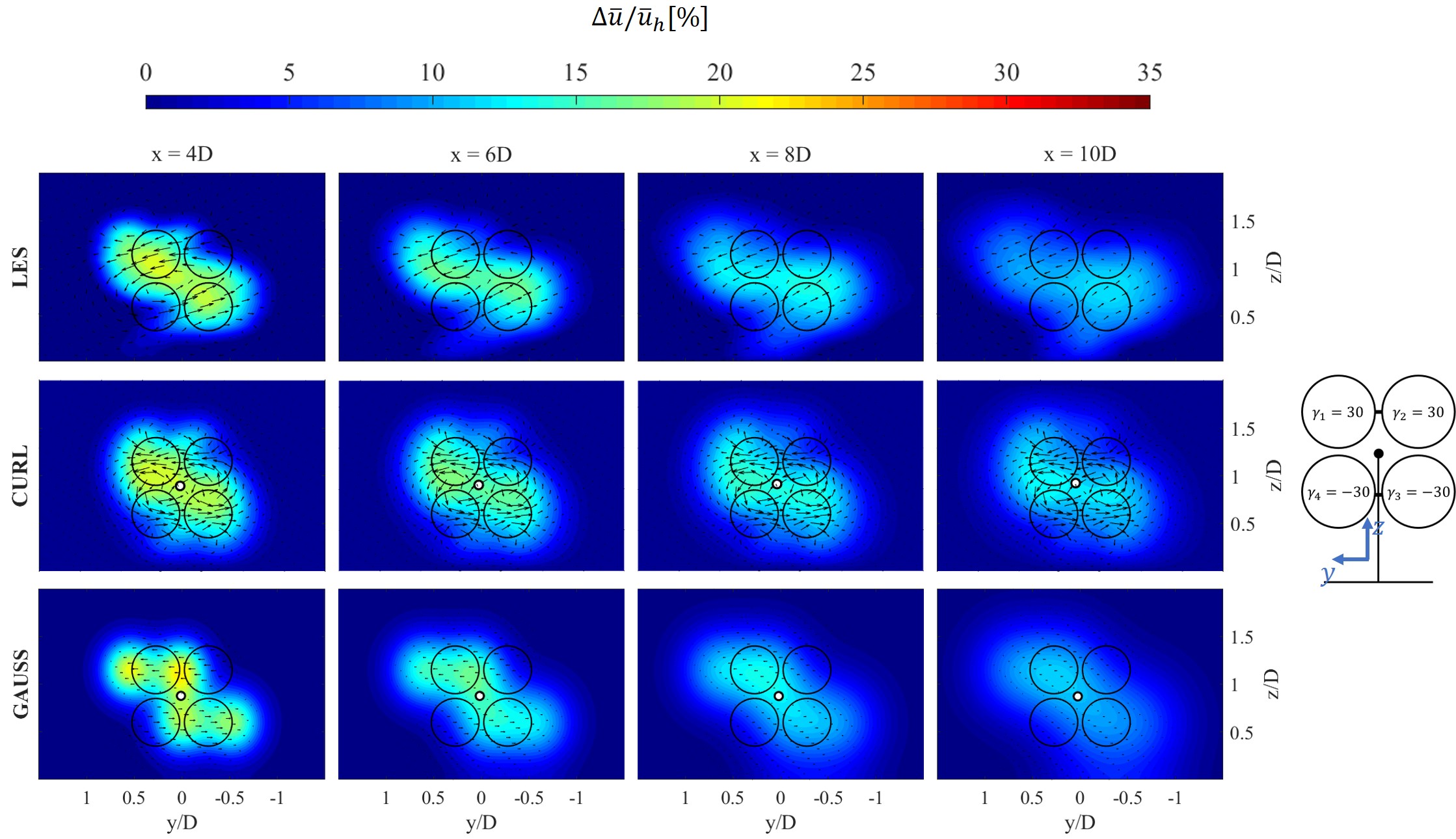}
  \caption{Crossed rotor wake cross sections at four downwind distances. Contours show normalised velocity deficit, $\Delta \bar{u}/\bar{u}_h$, and vectors indicate cross-stream velocity. Black circles illustrate rotor swept area, and a white dot denotes the wake centroid.}
\label{fig:div_lat_veldef}
%\vspace*{-5mm}
\end{figure*}

The first stage of analysis was to qualitatively inspect the wakes generated by each yaw configuration. Hence, cross sections of the velocity field at streamwise distances of $4D,6D,8D,$ and $10D$ were examined for each yaw configuration, as shown in Figs. \ref{fig:z_yaw_veldef}-\ref{fig:div_lat_veldef}. Contours of velocity deficit are illustrated in each case, with vectors of cross-stream velocity superposed. The centroid is illustrated by the white dot, the calculation of which is explained in \S\ref{sec:yc_Sy}, and the swept area of rotors is shown with black circles. Results from each of the three models are displayed in successive rows. LES predictions are used as a reference with which lower-fidelity curled-wake and Gaussian models can be compared.

\black The three models agree well in general prediction of wake behaviour, though there are some notable differences in deflection magnitudes and the magnitude of velocity deficit that will be qualitatively discussed in the following section. Quantitative comparison of model predictions is covered in more depth in \S \ref{sec:yc_Sy}, where wake widths and deflections are characterised mathematically. \black

Examining all configurations together, it is clear that the velocity deficit is lower in all yawed arrangements when compared with the baseline, a result of lower thrust forces exerted on the yawed rotors. As expected, the divergent and crossed rotor cases (Figs. \ref{fig:div_ver_veldef} and \ref{fig:div_lat_veldef}) appear to have the lowest velocity deficits since individual rotor wakes are directed in opposite lateral directions, thereby increasing wake expansion rates. In these configurations, rotor wakes also remain distinct over a larger streamwise distance, whereas rotor wakes interact and overlap faster in the baseline, equal yaw, and especially in convergent cases. Differences between configurations appear largest at short downwind distances, both in terms of magnitude and shape; velocity deficit levels further downstream are more similar between cases and wake boundaries are not so sharply defined.

Another pattern common to all configurations is a higher velocity deficit and deflection in top rotors compared with bottom rotors. This can be seen for all arrangements, but perhaps most clearly at $x/D = 4$ for divergent rotors in Fig. \ref{fig:div_ver_veldef}, where the wake forms a butterfly-like shape. A possible explanation for this may be offered by higher thrust forces exerted on top rotors, compounded by lower turbulence levels at greater heights. Since velocity increases with height due to the simulated atmospheric boundary layer, there will be a greater thrust force developed on top rotors, which will in turn lead to larger velocity deficits by conservation of momentum. Deflection has also been shown to be influenced by thrust force, as well as by turbulence intensity. \citet{jimenez2010application} found that a higher thrust leads to a greater deflection and \citet{bastankhah2016experimental} suggested that lower turbulence intensities lead to larger wake deflection. Both results corresponded to yaw of single-rotor turbines, however, it appears that similar relationships are present in multirotor arrangements.

Examining the baseline case (Fig. \ref{fig:z_yaw_veldef}), the most notable feature is how the velocity deficit region begins as a square array of individual rotor wakes at $x/D = 4$, before merging to a more circular shaped single wake as they move downstream to $x/D = 10$. Such behaviour verifies previous findings by \citet{van2019power} and \citet{bastankhah2019multirotor}, in which similar wake transitions were observed. In this arrangement, cross-stream velocities are small and appear to have no distinct pattern. \black The results of the curled-wake model are not in agreement with the LES data at short downwind distances, while the agreement improves further downstream. It seems that the curled-wake model over-predicts flow mixing such that rotor wakes already form a single wake at $x=4D$, which is not in agreement with the LES data. The Gaussian model predictions agree better with the LES data in this case, with only small disparities in terms of velocity deficit. \black

For equally yawed rotors, Fig. \ref{fig:eq_yaw_veldef} shows how all rotor wakes are deflected in one direction. Velocity deficit is lower than the baseline and the wake appears to span wider across the domain, indicating some potential for reduced wake losses. Moreover, formation of a kidney-shaped (curled) wake cross-section can be identified as the wake moves downstream, a phenomenon observed in similar studies of yawed single-rotor turbines \citep{bastankhah2016experimental,howland2016wake}. This feature is most likely the result of the counter-rotating vortex pair (CVP) that is also clearly present, again a typical characteristic of yawed single-rotor turbines. It would be expected that there would be CVPs associated with each rotor, however, in the LES it appears that some vortices merge or cancel out, leaving only one vortex pair. \black Lower-fidelity models appear less able to capture this merging, where CVPs remain distinct in the curled-wake model, and are not present in the Gaussian model. \black

The results of the equal yaw configuration also highlight the utility of visually inspecting wake cross sections in addition to mathematical characterisations of wake properties. In comparison to single-rotor turbines, multirotor turbines are capable of producing quite different wakes depending on rotor yaw arrangement. In this case, there is a significant difference between the overall lateral width across the domain and the width at hub height. While such complex wake distributions can offer significant advantages in reducing wind plant losses, it does mean that mathematical characterisations are less effective in fully describing the nature of the velocity deficit distribution. Hence, clear representation and close inspection of the flow field is a useful complement to quantitative analysis when developing an understanding of yawed multirotor wakes.

In the divergent rotor configuration, rotor wakes are deflected laterally outward, as shown in Fig. \ref{fig:div_ver_veldef}. This leads to the formation of a butterfly-shaped wake at $x/D = 4$ which transforms to more of a `V'-shape at $x/D = 10$. The velocity deficit is much lower than the baseline at all streamwise distances, indicating potential for reduced wake losses. The opposite effect can be seen for convergent rotors in Fig. \ref{fig:con_ver_veldef}, where rotor wakes are directed toward the lateral centre ($y = 0$). This leads to formation of a narrow wake, which widens at the base due to the ground effect. Though rotor wakes add up in this case, the velocity deficit is still lower than the baseline case. In both divergent and convergent arrangements, there appear to be four primary vortices of in-plane velocity, where direction of rotation depends on whether deflection is positive or negative. Therefore, vortex rotation directions are opposite in Figs. \ref{fig:div_ver_veldef} and \ref{fig:con_ver_veldef}. \black Again, individual rotor CVPs are observed in the curled-wake model but not in the LES data.

In these two configurations, deflection appears to be over-predicted by the Gaussian model, leading to two distinct wakes in the divergent case, and a high velocity deficit single wake in the convergent case. This can be explained by the fact that in these two configurations yawed rotors on each side of the turbine induce lateral velocities in opposite directions. As a result, their wake deflection, particularly in the convergent case, is expected to be less than an isolated rotor.  However, the Gaussian model is not able to capture this interaction between rotor wakes. In these configurations, the curled-wake model is able to provide more realistic predictions as it solves governing flow equations for all rotors at the same time. \black

Finally, in the crossed rotor arrangement (Fig. \ref{fig:div_lat_veldef}), top rotor wakes are directed in the opposite direction to bottom rotor wakes. This forms the wake into a wide asymmetric shape spanning a large lateral distance. Like the divergent case, velocity deficit is much lower than the baseline, again making this a promising option for increasing downstream power outputs. \black The Gaussian model captures this well; however, the deflection and velocity deficit levels of the curled-wake model are under-predicted. As before, it appears that individual CVPs combine to form fewer large vortices. In the LES visualisation, three cross-stream vortices are identifiable---one central and two smaller instances in the top left and bottom right of the domain. \black

Overall, it is clear that the divergent and crossed rotor configurations produce the greatest wake expansion, which may suggest that these configurations will be the most effective in minimising wake losses. However, the wake deflection caused by the equal yaw case is effective at reducing the velocity deficit within the planform area of the rotors so it may also be expected to perform well. Furthermore, the wake channelling displayed in the convergent arrangement may find its use in wind plant control and optimisation, depending on the arrangement of downwind turbines. For example, in a staggered wind farm layout where neither expansion nor steering are likely to be useful, channelling can guide wakes between downstream turbines.

\begin{figure*}[t!] %***u_x***
\centering
%\vspace*{-10mm}
  \begin{subfigure}[t]{0.8\textwidth}
    \includegraphics[width=\textwidth]{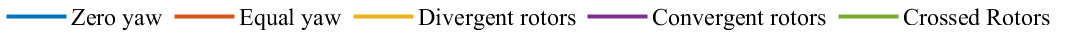} 		 	
    \vspace*{-10mm}
  \end{subfigure}

  \begin{subfigure}[t]{0.48\textwidth}
  \includegraphics[width=\textwidth]{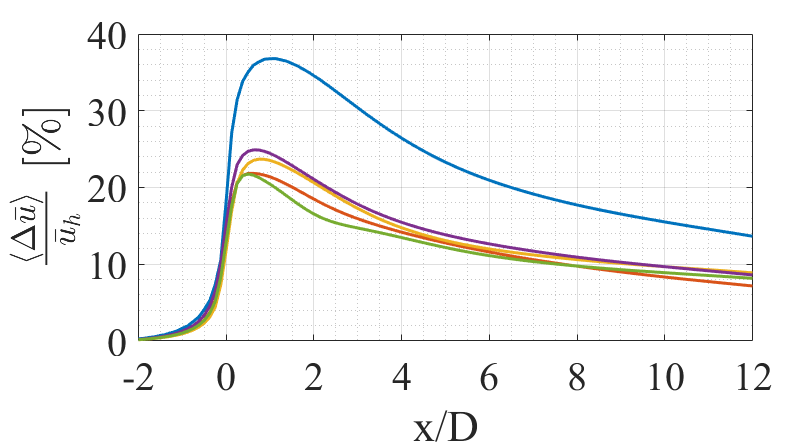}
    \caption{Variation of normalised velocity deficit averaged over the rotor area, $\langle\Delta \bar{u}\rangle/\bar{u_h}$, with downstream distance, $x/D$.}
    \label{fig:x_veldef}
  \end{subfigure}
  ~
    \begin{subfigure}[t]{0.48\textwidth}
  \includegraphics[width=\textwidth]{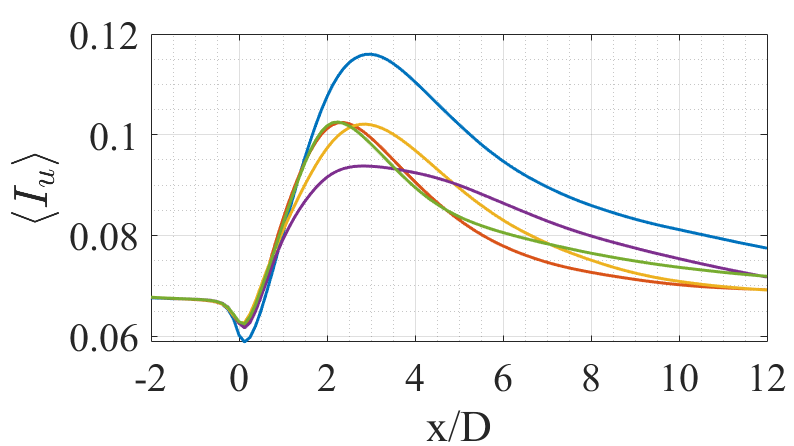}
    \caption{Variation of turbulence intensity averaged over the rotor area, $\langle I_u \rangle$, with downstream distance, $x/D$.}
    \label{fig:x_Iu}
  \end{subfigure}
  \caption{Variation of rotor-averaged velocity deficit and turbulence intensity with downstream distance.}
  \label{fig:x_Iu_veldef}

\includegraphics[width=\textwidth]{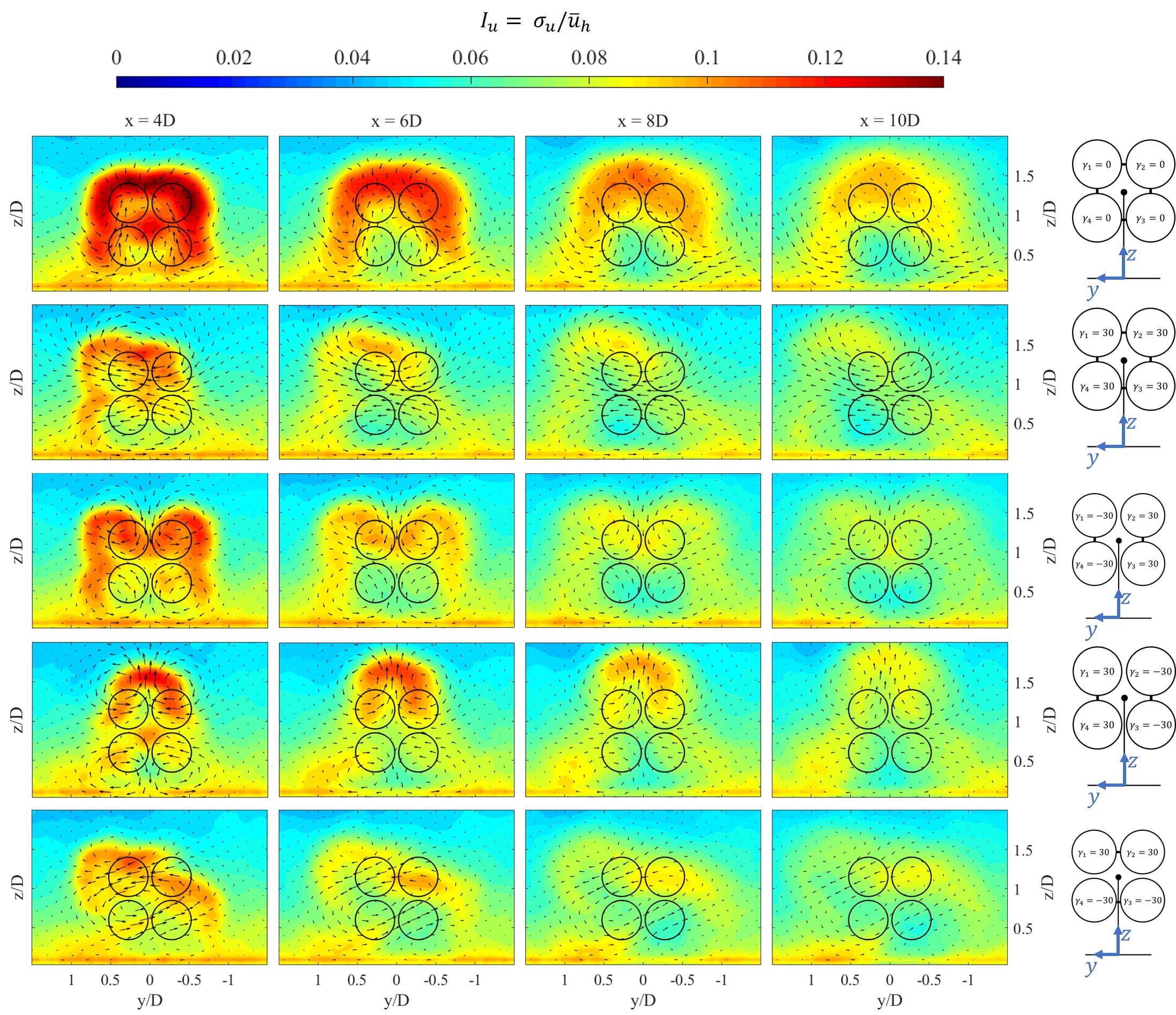}
  \caption{Cross sections of turbulence intensity for all configurations at four downwind distances. Contours show turbulence intensity, $I_u$, vectors indicate cross-stream velocity, and black circles illustrate rotor swept area.  Configuration schematics are shown to the right of the plots to which they correspond.}
\label{fig:turb_int}
\vspace*{-10mm}
\end{figure*}

The development of the wakes can be further understood by examining how the normalised velocity deficit averaged over the rotor area, $\langle\Delta \bar{u}\rangle/\bar{u_h}$, varies with downwind distance, as illustrated in Fig. \ref{fig:x_veldef}. The results of this analysis agree broadly with the velocity deficit contours in Figs. \ref{fig:z_yaw_veldef}--\ref{fig:div_lat_veldef} in that, the baseline case exhibits the highest velocity deficit levels, followed by the convergent case. However, the divergent case appears to cause only slightly lower velocity deficit levels than the convergent case, and in the far wake, the two become approximately equal. The crossed rotor configuration initially has the lowest levels of velocity deficit, which remains the case up to approximately $8D$ downstream, beyond which the equal yaw configuration performs better. This may be explained by the initially fast wake recovery rate of the crossed rotor case, which then slows further downstream. Nevertheless, in the region of interest ($5D-7D$, where a second turbine may be placed), the crossed rotor arrangement still has the lowest velocity deficit, on average.

Following the above analysis of velocity deficit, turbulence intensity, $I_u$, was investigated in a similar manner as illustrated in Figs. \ref{fig:x_Iu} and \ref{fig:turb_int}. Quantitative data is presented in Fig. \ref{fig:turb_int} which displays contours of turbulence intensity, $I_u$, for all yaw configurations at downwind distances of $4D$, $6D$, $8D$, and $10D$. In this case, only LES data are displayed since only this model is able to predict turbulence intensity. Overall, the figure shows that, compared with the baseline case, yawing rotors significantly reduce the turbulence intensity in the wake. This is due to the fact that the wake is weaker for yawed rotors, with less mean flow shear. Crucially, lower turbulence levels may present benefits for downwind turbines in terms of lower flow-induced unsteady structural loads.

One notable behaviour displayed here is that the distribution of turbulence appears to be affected by cross-stream velocity components, as seen most clearly in convergent and divergent configurations. In the convergent configuration, the position of maximum turbulence intensity is pushed upward as the flow develops downstream, a movement which clearly follows the direction of the cross-stream flows that are generated by the rotor CVPs. Similarly, the turbulent flows in the divergent configuration also appear to follow the associated cross-stream velocity. The effects of this may be seen in Fig. \ref{fig:x_Iu}, which displays variation in turbulence intensity averaged over rotor area, $\langle I_u \rangle$, with downwind distance. Due to the cross-stream motions induced by the presence of CVP, the turbulence intensity is less for the convergent case in the near wake. However, since the wake recovery is slow for this case, the turbulence intensity ultimately becomes larger than other cases at large downwind distances. The most consistently low rotor-averaged turbulence intensity is associated with the equal yaw case, which may therefore be expected to deliver the most benefits in terms of reducing fatigue on downwind turbines. Further quantitative analysis is carried out in the following section, in which wake widths and deflections are mathematically characterised.

\clearpage

\subsection{Centroid and Wake Width} \label{sec:yc_Sy}

Following qualitative examination of wake cross sections, a quantitative analysis was carried out to characterise wake centroids and widths. The centroid was calculated using an arithmetic mean of velocity deficit values within a given streamwise plane. The integration domain considered was the same as that shown in Figs. \ref{fig:z_yaw_veldef}--\ref{fig:div_lat_veldef}, which is large enough to ensure that velocity deficit becomes zero at the boundary. In this respect, the analysis is analogous to a centre of mass calculation, with the lateral centroid location from the turbine centre given by

\begin{equation} \label{eq:cent}
y_c = \frac{\int y \Delta \bar{u} dA}{\int \Delta \bar{u} dA}.
\end{equation}
A similar calculation may be performed to find the vertical centroid location, though this typically remains close to hub height. The lateral width of the total multirotor wake can be represented by its standard deviation, $\sigma_y$, given by

\begin{equation}\label{eq:sigma}
\sigma_y = \sqrt{\frac{\int (y - y_c)^2 \Delta \bar{u} dA}{\int \Delta \bar{u} dA}}.
\end{equation}
\black Although Equations. \ref{eq:cent} and \ref{eq:sigma} should be computed numerically for the LES data as well as curled-wake predictions, analytical relationships can be found for the Gaussian wake model. Derivations for these are presented in \ref{app:der} and results are repeated here for convenience.

\begin{equation} \label{eq:an_cent}
  y_c =
  \begin{cases}
	0 & \text{if equal and opposite}, \\
	\delta_1 & \text{if exactly equal},
  \end{cases}
\end{equation}
\begin{equation} \label{eq:an_width}
  \sigma_y =
  \begin{cases}
	 \sqrt{y_1^2 + \sigma_{y_1}^2} & \text{if $\gamma_1 = \gamma_2 = \gamma_3 = \gamma_4$ \qquad (equal/zero yaws)}, \\
	 
	\sqrt{y_1^2 + \sigma_{y_1}^2 + \delta_1^2} & \text{if $\gamma_1 = \gamma_2 = -\gamma_3 = -\gamma_4 $ \; (crossed rotors)}, \\
	
	\sqrt{y_1^2 + \sigma_{y_1}^2 + \delta_1^2 + 2\delta_1y_1} & \text{if $\gamma_1 = -\gamma_2 = -\gamma_3 = \gamma_4 $ \; (divergent/convergent rotors)},
  \end{cases}
\end{equation}
where $\delta_1$, $y_1$, and $\sigma_{y_1}$ represent the deflection, lateral offset, and lateral wake width of $R1$, respectively. \black

\begin{figure*}[t] %***CENT***
\centering
  \begin{subfigure}[t]{0.8\textwidth}
    \includegraphics[width=\textwidth]{legend.png}
    \label{fig:legend}
    \vspace*{-5mm}
  \end{subfigure}
  ~
    \begin{subfigure}[t]{0.48\textwidth}
    \includegraphics[width=\textwidth]{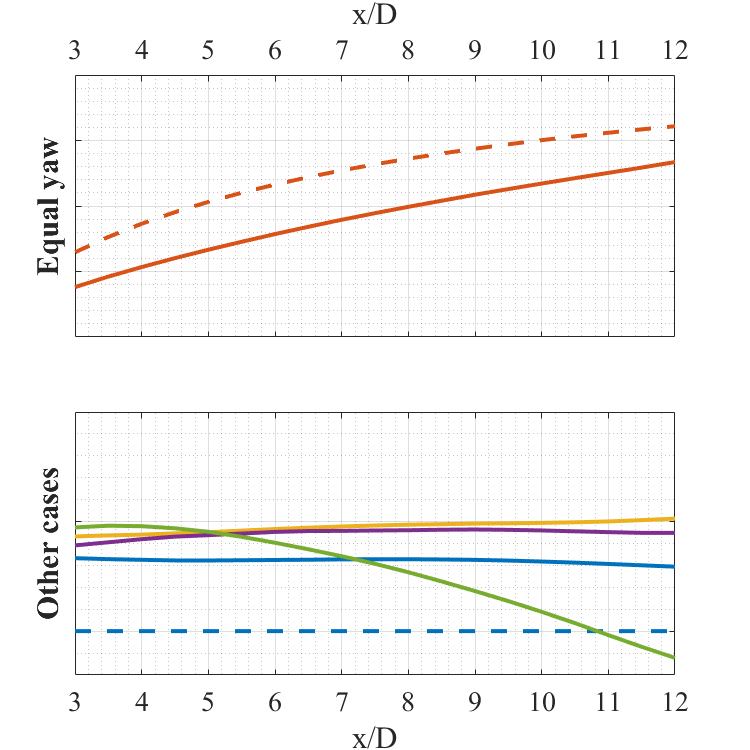}  
    \caption{Gaussian model versus LES. \blue All Gaussian predictions apart from the equal yaw case collapse to $y_c/D = 0$. \black}
    \label{fig:gauss-LES_yc_comp}
  \end{subfigure}
  ~
  \begin{subfigure}[t]{0.48\textwidth}
    \includegraphics[width=\textwidth]{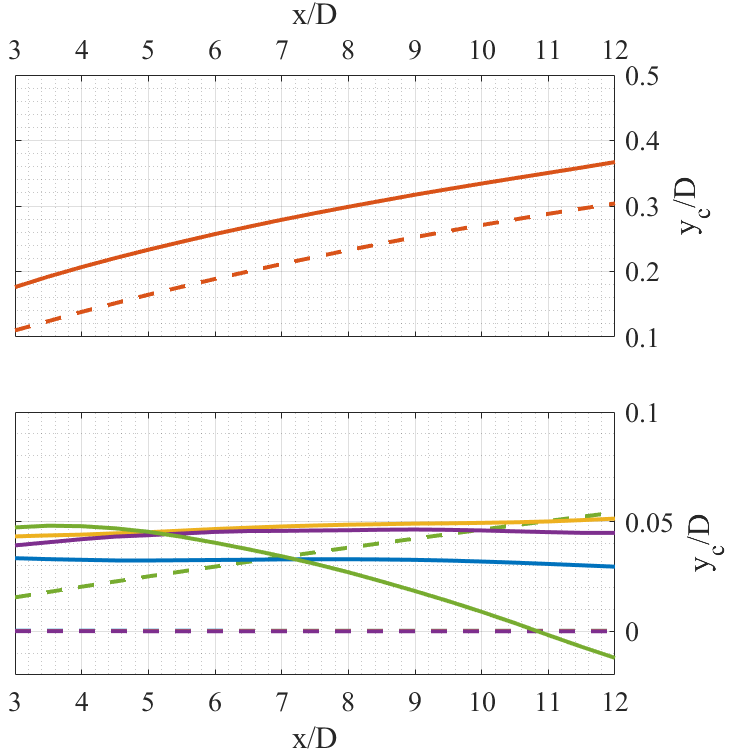}
    \caption{Curled-wake model versus LES. \blue All curled-wake predictions apart from the equal yaw and crossed rotors collapse to $y_c/D = 0$. \black}
    \label{fig:cw-LES_yc_comp}
  \end{subfigure}
\caption{Lateral centroid deviation, $y_c/D$, with streamwise distance, $x/D$, for all five yaw configurations. Solid lines denote LES and dashed lines denote Gaussian and curled-wake predictions. \blue Results for the equal yaw case are plotted separately since this data falls within a different range to the other cases. \black}
\label{fig:cent}
\end{figure*}  

\begin{figure*}[t] %***WIDTH***
\centering  
  \begin{subfigure}[t]{0.8\textwidth}
    \includegraphics[width=\textwidth]{legend.png}
    \vspace*{-10mm}
  \end{subfigure}
  \begin{subfigure}[t]{0.48\textwidth}
  \includegraphics[width=\textwidth]{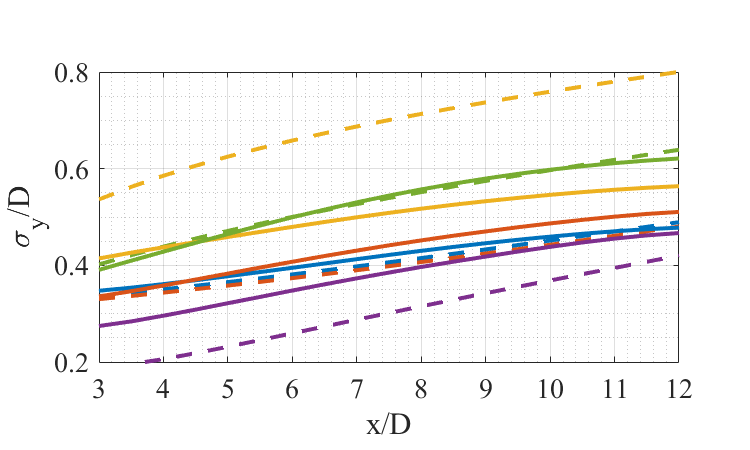}  
    \caption{Gaussian model versus LES}
    \label{fig:gauss-LES_Sy_comp}
  \end{subfigure}
  ~
  \begin{subfigure}[t]{0.48\textwidth}
  \includegraphics[width=\textwidth]{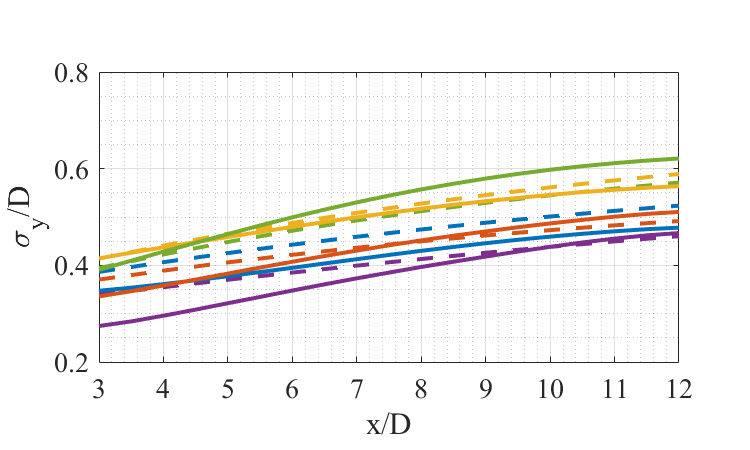}
    \caption{Curled-wake model versus LES}
    \label{fig:cw-LES_Sy_comp}
  \end{subfigure}
  \caption{Wake width, $\sigma_y/D$, with streamwise distance, $x/D$, for all five yaw configurations. Solid lines denote LES and dashed lines denote Gaussian and curled-wake predictions.}
    \label{fig:Sy}
\end{figure*}

The wake centroid locations of all configurations are plotted over a streamwise distance range $3D-12D$ in Fig. \ref{fig:cent}. LES results are plotted with solid lines and dashed lines denote Gaussian and curled-wake solutions in Figs. \ref{fig:gauss-LES_yc_comp} and \ref{fig:cw-LES_yc_comp}, respectively. \blue Results for the equal yaw case are plotted separately since this data falls within a different range to the other cases. For this arrangement, all rotor wakes are deflected in the same lateral direction resulting in a large overall turbine wake deflection. For all other cases, LES data suggests that that wakes remains close to the lateral centre ($y = 0$). \black These small overall wake deflections are a result of most configurations generating two sets of rotor wakes with equal and opposite deflections. However, while most cases remain close to $y = 0$, none are centred exactly at this location with most exhibiting a positive lateral offset. The wake of the crossed rotors also shows some negative centroid movement after \blue $x/D = 5$ \black. 

Fig. \ref{fig:gauss-LES_yc_comp} shows how these patterns are captured by the Gaussian model, though deflection in the equal yaw configuration is over-predicted. Between $x/D = 3$ and $x/D = 12$, the error between LES and analytical predictions for this case ranges between $0.053D$ and $0.076D$, with the maximum error occurring at approximately $x/D = 6$. The analytical model is also unable to capture the positive offset of other configurations, for which centroid locations are all approximated to $y = 0$. \blue Note that the lower plots show a much finer scale in the y-axis, exaggerating deviations between LES and other models. Extensions of the Gaussian model for multiple yawed wakes have been developed \citep[e.g.,][]{zong2020momentum}, but these models have been outside of the scope of this investigation. Similar to the Gaussian model, the curled-wake model predicts the centroid location at or close to lateral centre for most yawed cases, as shown in Fig. \ref{fig:cw-LES_yc_comp}. For the crossed rotor case, a positive centroid offset is predicted, however the negative gradient is not captured. Under equal yaw conditions the centroid variation is under-predicted with an approximately constant error of $0.068D$ over the streamwise range. \black

LES wake widths are plotted over the same streamwise range for all configurations in Fig. \ref{fig:Sy}, with the Gaussian and curled-wake results superposed. The LES data show that wake widths are very similar for zero and equal yaw arrangements, while divergent and crossed rotor cases exhibit a significantly higher wake expansion. The convergent case exhibits a narrower wake, though this is less pronounced at greater downstream locations. Comparing these results with Figs. \ref{fig:z_yaw_veldef}--\ref{fig:div_lat_veldef} indicates that $\sigma_y$ provides an effective measure of the wake widths. For example, examining Figs. \ref{fig:z_yaw_veldef} and \ref{fig:eq_yaw_veldef}, it can be seen that wakes widths are in fact similar in spite of their different wake shapes. While the equal yaw wake is narrower at hub height, the overall lateral spread of velocity deficit is very close to that of the baseline. The large widths generated by divergent and crossed rotors are also captured by $\sigma_y$, as is the narrow velocity deficit of the convergent case.

\black The Gaussian model acceptably captures most of this behaviour; however, there is significant error in modelling of divergent and convergent rotor arrangements. The width of the wake from the convergent rotor case is under-predicted, leading to errors between $0.047D$ and $0.086D$ over the plotted streamwise range, with larger errors in the near wake. By contrast, the divergent case is over-predicted by the analytical model, leading to errors between $0.12D$ and $0.24D$ over the plotted streamwise range, with larger errors in the far wake. This lines up with what was seen in the velocity deficit contours, where rotor wakes were deflected either too far away or too far toward the lateral centre. It is likely that this is caused by the magnitude of the final term in Equation. \ref{eq:an_width}, ($2\delta_1y_1$), which may have a disproportionately large effect on $\sigma_y$. The analytical model also estimates the equal yaw wake to be narrower than the baseline since individual rotor wakes are narrower under yawed conditions. However, this is not seen in the LES results, indicating that the effect of individual rotor wake widths does not have a significant impact on the overall wake of the multirotor turbine. The curled-wake model displays the same patterns as the Gaussian solutions predicting convergent and equal yaw cases to be narrower than the baseline, while divergent rotors and crossed rotors are wider. However, the spread of the data is much smaller and hence the different cases are less distinguishable by their width. Notably, this leads to over-predictions of the convergent rotor wake width and under-prediction of the crossed rotor width. Errors in the convergent case are between $0$ and $0.065D$ over the plotted streamwise range, while errors in the crossed rotor case over the same range are between $0.004D$ and $0.05D$. That is, for the convergent case, curled-wake predictions are closer to the LES in the far wake, whereas for the crossed case, predictions are closer to the LES data in the near wake. \black

The quantitative characterisations presented in this section confirm the analysis of the velocity deficit contours given in Figs. \ref{fig:z_yaw_veldef}--\ref{fig:div_lat_veldef} in that the divergent and crossed rotors produce the largest wake expansion. The widths of the baseline and equal yaw cases are approximately similar, while the convergent case facilitates some wake channelling. It should be acknowledged, however, that the wake deflection caused by the equal yaw arrangement may also be effective in reducing wake losses in spite of its lower wake expansion. Further insights into the overall utility of each configuration are provided in the following section through an analysis of power output.

\subsection{Power Analysis} \label{sec:power}

\begin{figure*}[t] %***POWER***
\centering
  \begin{subfigure}[t]{\textwidth}
  \centering
    \includegraphics[width=0.7\textwidth]{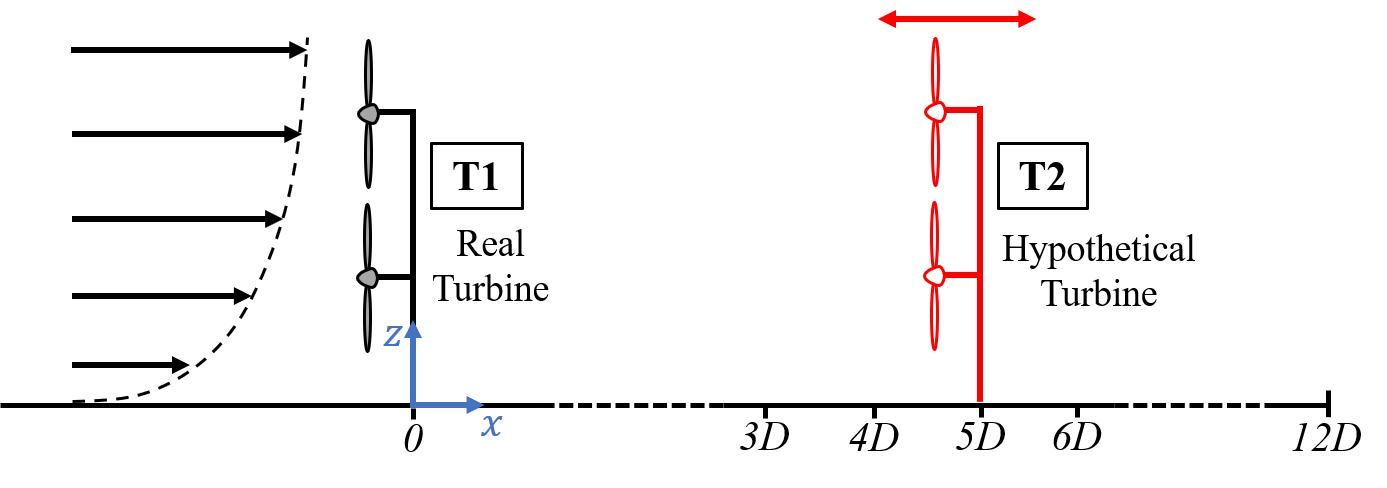}
    \caption{The two-turbine system used for power analysis, viewed from one side. Black arrows indicate the direction of incoming flow. The downstream turbine, T2, is rendered in red to indicate that it is hypothetical and movable between $x/D = 3$ and $x/D = 12$.}
    \label{fig:power_sch}
  \end{subfigure}
  \vspace*{3mm}
  
  \begin{subfigure}[t]{0.8\textwidth}
    \includegraphics[width=\textwidth]{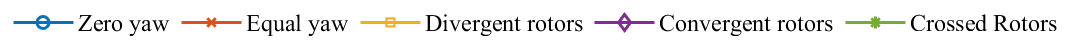} \label{fig:power_leg}
  \end{subfigure}
  \vspace*{-3mm}

  \begin{subfigure}[t]{0.48\textwidth}
    \includegraphics[width=\textwidth]{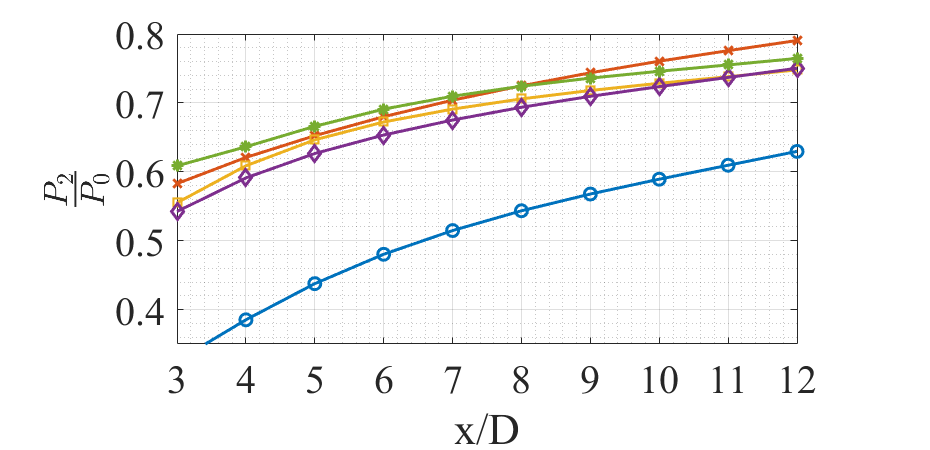}  
    \caption{Power output of the downstream turbine, $P_2$, normalised by the power output of the upstream turbine at zero yaw, $P_0$.}
    \label{fig:P2}
  \end{subfigure}
  ~
  \begin{subfigure}[t]{0.48\textwidth}
    \includegraphics[width=\textwidth]{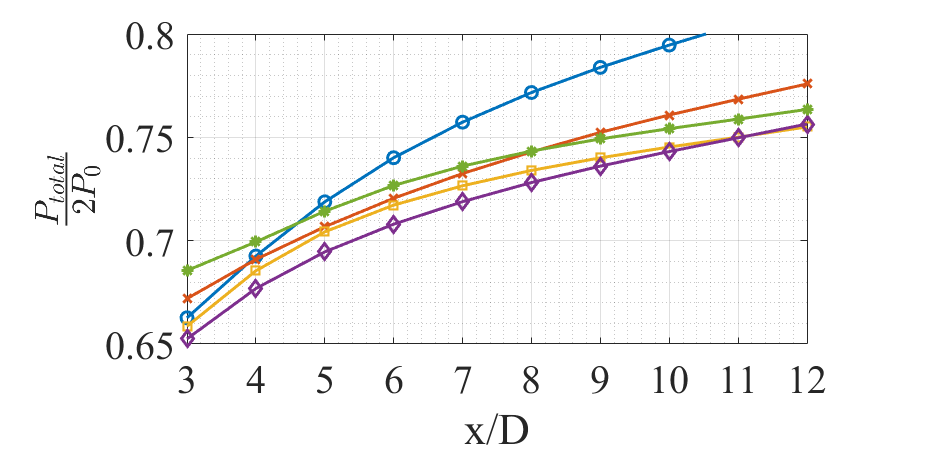}    
    \caption{Total power output of both the upstream and downstream turbines, normalised by twice the power output of the upstream turbine at zero yaw, $2P_0$.}
    \label{fig:Ptot}
  \end{subfigure}
\caption{Power output of the downstream turbine and both turbines for the five different yaw configurations.}
\label{fig:power}
\end{figure*}

Finally, the yaw configurations were evaluated by their effect on the power production of two turbines: one downstream of the other as illustrated in Fig. \ref{fig:power_sch}. We do this by using the LES data and placing a hypothetical turbine at different downstream locations.
The power of the upstream turbine, $P_1$, was calculated using Equation. \ref{eq:power}. Values of the incoming velocity, $\bar{u}_{\infty}$, were taken at a location upstream such that they were not affected by the presence of the rotor. The normalised power, $P/P_0$, was 76\% for yawed rotors, as presented in Fig. \ref{fig:PP0}. The power of the second hypothetical downstream turbine, $P_2$, was calculated in a similar manner; however, since the downstream turbine is not modelled in the LES, the values of the incoming velocity for the second turbine could be taken at the exact rotor location. Note that for all cases, the downstream turbine is kept unyawed.

The results of this analysis are presented in Figs. \ref{fig:P2} and \ref{fig:Ptot} respectively, which show the downstream turbine output, $P_2$, and the total power output, $P_{total}$, for a range of downstream turbine placements between $x/D = 3$ and $x/D = 12$. The power of the upstream turbine under zero yaw, $P_0$, was used to normalise the data. As expected from previous analysis, Fig. \ref{fig:P2} shows downstream power output to be significantly higher when the upstream turbine is yawed; a result of the lower velocity deficit associated with yawed rotors. $P_2$ is initially highest for the crossed rotor case, most likely due to its high wake expansion in the near wake. However, beyond $x/D = 8$, the equal yaw case leads to a greater power output from the downstream turbine, which may be caused by the large overall wake deflection caused by this configuration. The divergent arrangement produces slightly more power than the convergent case; however this difference is negligible in the far wake. Notably, this follows the same pattern as that of the averaged velocity deficit in Fig. \ref{fig:x_veldef}; as would be expected, there is a strong link between velocity deficit levels and power output from downstream turbines.

The combined power output of both turbines is plotted in Fig. \ref{fig:Ptot}, which shows higher power output from some yawed configurations in the near wake up to $x/D = 5$. Though benefits from yawed configurations are modest, it should be noted that power gains would be expected to be much more pronounced for a larger wind farm. In a typical wind farm context it is likely that power gains would not only be seen in the second turbine row, but in turbines much further downstream, as shown in prior studies \citep{bastankhah2019wind}. Such power gains are largely a result of ``secondary steering'', in which high flow entrainment caused by the CVPs of yawed rotors is compounded over several rows of turbines \cite{King2020,fleming2018simulation}. Furthermore, if wake-steering strategies are taken into account in the wind farm design process, turbines may be more closely spaced, meaning that the near wake power gains observed in Fig. \ref{fig:Ptot} could be more relevant. Finally, power gains may also vary for different inflow conditions, such as the level of incoming turbulence, $I_0$. It is expected that application of rotor yaw with a lower value of $I_0$ would be much more effective at accelerating wake recovery and delivering power gains.

Overall, the results of this power analysis are in agreement with the previous data. Among different yawed configurations, the crossed rotor case appears to perform best in the region of interest where downstream turbines are most likely to be placed, while the equal yaw case produces more power in the far wake. The divergent arrangement delivers more power overall than the convergent case; however, in the far wake the difference between them is minimal. Power gains are seen in the near wake; however, it is acknowledged that this two-turbine system does not reflect the larger power gains that might be expected in a wind farm or a multiple turbine arrangement.

\section{Conclusions}

In this paper, wake steering was combined with the concept of a multirotor turbine to extend wind farm control capabilities. Large-eddy simulations, curled-wake, and Gaussian modelling approaches were used to test the effectiveness of applying wake-steering methods to multirotor turbines. A range of five different rotor yaw configurations (including a reference case) was investigated by closely examining the wakes at various downstream locations. A qualitative method was employed first, in which wake cross sections were examined in terms of their distribution and magnitude of velocity deficit. A key finding from this analysis was that the divergent and crossed rotor configurations were able to produce a significantly larger wake expansion than the baseline, and also generated much lower velocity deficits. The other two arrangements (equal yaw and convergent) were able to channel and redirect the wake, which may also be used to enhance wake recovery/redirection. A subsequent examination of rotor-averaged velocity deficit confirmed that the high wake expansion of the crossed rotor case was most effective at moderate downwind distances, whereas the large deflection from the equal yaw arrangement brought about the lowest velocity deficit at far downwind distances. Overall, higher wake deflection was observed in top-row rotors, which corresponds to their higher thrust forces as well as their lower incoming turbulence intensities, confirming previous findings from single-rotor studies. Cross-stream velocity components were also given some attention since they can have a significant effect on how the wake develops as it moves downstream. Counter-rotating vortex pairs were identified at each rotor which, in many cases, cancelled or combined with each other to form larger vortices.

A mathematical characterisation of wake widths and deflections was subsequently performed, which largely confirmed the findings of the cross-sectional wake analysis. The centroid calculations appear to agree well with what can be seen by inspection of the velocity field and acceptably predict the overall turbine wake movement. Similarly, the wake width variation along the streamwise range was in good agreement with what was shown in the wake cross sections. Finally, an analysis of power production was carried out as a concluding assessment of the multirotor yaw scheme for a case with two multirotor turbines with various inter-turbine spacing. Modest power gains were observed for crossed and equal configurations, which are expected to be more pronounced in a wind farm context.

The two lower-fidelity models showed an acceptable agreement with the LES data; however, some discrepancies were observed. \blue Extensions to the Gaussian model could be made to improve its overall fit with LES results. Further work could be aimed at reducing the over-prediction of wake deflection that was observed in some yaw arrangements, most notably in convergent and divergent configurations. This could be achieved through more detailed modelling of cross-stream velocity components and CVP interactions. Similarly, the curled-wake model needs some tuning to correct the expansion rate. Improvements to curled-wake model results can be obtained by solving the transport equations in the spanwise and wall-normal directions. These improvements will be part of future work. \black

The effects of varying inflow conditions may also be investigated further, such as different wind shear, veer, and turbulence intensities, as well as different thrust coefficients, yaw magnitudes, and tip clearances. Also, to shed light on the effects of closely spaced adjacent rotors on the structure and dynamics of the near-wake region, high-resolution numerical simulations using actuator-line techniques and advanced vortex methods can be utilized. Finally, the effects of multirotor yaw schemes could be tested on a larger scale, such as in the context of a wind farm.

%\printendnotes

% Submissions are not required to reflect the precise reference formatting of the journal (use of italics, bold etc.), however it is important that all key elements of each reference are included.

\paragraph{Declaration of Competing Interest} This work was authored in part by the National Renewable Energy Laboratory, operated by Alliance for Sustainable Energy, LLC, for the U.S. Department of Energy (DOE) under Contract No. DE-AC36-08GO28308. Funding provided by the U.S. Department of Energy Office of Energy Efficiency and Renewable Energy Wind Energy Technologies Office. The views expressed in the article do not necessarily represent the views of the DOE or the U.S. Government. The U.S. Government retains and the publisher, by accepting the article for publication, acknowledges that the U.S. Government retains a nonexclusive, paid-up, irrevocable, worldwide license to publish or reproduce the published form of this work, or allow others to do so, for U.S. Government purposes.

\paragraph{Data Availability} %Data may be available by contacting the corresponding author.
The data that support the findings of this study are available from the corresponding author upon reasonable request.

\appendix
\black
\section{Gaussian analytical model derivation for wake centroid and width} \label{app:der}    %% Appendix A

The objective of this appendix is to clarify the derivation of two key wakes parameters, namely the wake centroid and wake width, based on the Gaussian multirotor wake model. First, the integrals of Equations. \ref{eq:cent} and \ref{eq:sigma} are analytically computed, and subsequently simplified using assumptions of turbine geometry and wake symmetry. This facilitates direct calculation of the centroid location and wake width and circumvents the need to generate a velocity field.

\subsection{Wake Centroid Prediction}

The centroid of the wake generated by the multirotor turbine may be given by

\begin{equation} \label{eq:app_cent}
y_c = \frac{\int_{-\infty}^{\infty} y \Delta \bar{u} dA}{\int_{-\infty}^{\infty} \Delta \bar{u} dA} =  \frac{\sum_{n=1}^{4} \int_{-\infty}^{\infty} y \Delta \bar{u}_n dA}{\sum_{n=1}^{4} \int_{-\infty}^{\infty} \Delta \bar{u}_n dA}.  
\end{equation}
Using the Gaussian profile for velocity deficit given in Equation. \ref{eq:sr_veldef}, the $n^{th}$ integral in the numerator of Equation. \ref{eq:app_cent} may be evaluated 

\begin{equation}
\int_{-\infty}^{\infty} y \Delta \bar{u}_n dA = \int_{-\infty}^{\infty} \int_{-\infty}^{\infty} C_n y e^{-\frac{(y-y_n-\delta_n)^2}{2\sigma_{y_n}^2}} e^{-\frac{(z-z_h-z_n)^2}{2\sigma_{z_n}^2}} dy dz = 2 \pi C_n \sigma_{y_n} \sigma_{z_n}(y_n + \delta_n).
\end{equation}
Similarly, evaluating the $n^{th}$ integral in the denominator of Equation. \ref{eq:app_cent} gives 

\begin{equation} \label{eq:den}
\int_{-\infty}^{\infty} \Delta \bar{u}_n dA = \int_{-\infty}^{\infty} \int_{-\infty}^{\infty} C_n e^{-\frac{(y-y_n-\delta_n)^2}{2\sigma_{y_n}^2}} e^{-\frac{(z-z_h-z_n)^2}{2\sigma_{z_n}^2}} dy dz = 2 \pi C_n \sigma_{y_n} \sigma_{z_n}.
\end{equation}
The remaining terms in the numerator and denominator of Equation. \ref{eq:app_cent} are evaluated in a similar way, which allows the centroid to be written as

\begin{equation} \label{eq:A4}
y_c = \frac{\sum_{n=1}^{4} C_n \sigma_{y_n} \sigma_{z_n}(y_n + \delta_n)}{\sum_{n=1}^{4} C_n \sigma_{y_n} \sigma_{z_n}}.
\end{equation}

Next, we attempt to simplify Equation \ref{eq:A4} for the studied yaw configurations. Given the geometrical symmetry of a four-rotor turbine, the equation may be simplified by saying $y_1 = y_4 = -y_2 = -y_3 $. Moreover, rotors with the same yaw angle magnitude and thrust coefficient typically have similar values of maximum velocity deficit, $C_n$, and wake widths, $\sigma_{y_n}$ and $\sigma_{z_n}$. Therefore, Equation. \ref{eq:A4} may be simplified to

\begin{equation}
y_c = \frac{1}{4} \sum_{n=1}^4\delta_n.
\end{equation}
Deflection patterns for the studied cases fall into two categories: (1) where all deflections are equal and (2) where two deflections are equal and opposite to the other two. Hence, the centroid can be represented simply as

\begin{equation} \label{eq:app_yc_cases}
  y_c =
  \begin{cases}
	0 & \text{if equal and opposite deflection}, \\
	\delta_1 & \text{if exactly equal deflection}.
  \end{cases}
\end{equation}

\subsection{Wake Width Prediction}

The total wake width of the multirotor turbine can be represented by its standard deviation, $\sigma_y$. Here, for simplicity, the variance, $\sigma_y^2 $, is written. As before, this may be represented as the sum of contributions from each rotor

\begin{equation} \label{eq:app_var}
\sigma_y^2 = \frac{\int_{-\infty}^{\infty} (y - y_c)^2 \Delta \bar{u} dA}{\int_{-\infty}^{\infty} \Delta \bar{u} dA} = \frac{\sum_{n=1}^{4} \int_{-\infty}^{\infty} (y - y_c)^2 \Delta \bar{u}_n dA}{ \sum_{n=1}^{4} \int_{-\infty}^{\infty} \Delta \bar{u}_n dA}.
\end{equation}
The denominator is the same as in the case of the centroid (Equation. \ref{eq:den}). Evaluating the $n^{th}$ integral in the numerator of Equation. \ref{eq:app_var} gives

\begin{equation}
\int_{-\infty}^{\infty} (y - y_c)^2 \Delta \bar{u}_n dA= 2 \pi C_n \sigma_{y_n} \sigma_{z_n} \left(y_c^2 - 2y_c(y_n + \delta_n) + y_n^2 + 2\delta_n y_n + \sigma_{y_n}^2 + \delta_n^2 \right).
\end{equation}
If yaw magnitudes and wake widths are again assumed to be approximately equal then the $ 2 \pi C_n \sigma_{y_n} \sigma_{z_n}$ term may be cancelled from the top and bottom of the fraction in Equation. \ref{eq:app_var}, allowing the variance to be written as

\begin{equation}
\sigma_y^2 = y_c^2 + \frac{1}{4} \sum_{n=1}^{4} (y_n^2 + 2\delta_n y_n + \sigma_{y_n}^2 + \delta_n^2 - 2y_c(y_n + \delta_n)).
\end{equation}
The geometric simplifications used for the centroid may again be applied such that $y_1 = y_4 = -y_2 = -y_3$. Deflection magnitudes are also assumed equal, so the above equation can be expanded to

\begin{equation}
\sigma_y^2 = y_c^2 + y_1^2 + \sigma_{y_1}^2 + \delta_1^2 +  \frac{1}{2} \left[\delta_1(y_1 - y_c) - \delta_2(y_1 + y_c) - \delta_3(y_1 + y_c) + \delta_4(y_1 - y_c) \right].
\end{equation}
This relationship may be simplified for certain cases tested in this study. First, where all yaw angles and deflections are exactly equal, $ y_c = \delta_1 $. If the top rotors are yawed in an equal and opposite direction to the bottom rotors (crossed rotors) then $\delta_1 = \delta_2 = -\delta_3 = -\delta_4 $ and $ y_c = 0 $. Finally, if the left-side rotors are yawed in an equal and opposite direction to the right-side rotors (divergent/convergent rotors) then $\delta_1 = -\delta_2 = -\delta_3 = \delta_4 $ and again $ y_c = 0 $. Since $\delta_n \propto \gamma_n$, the standard deviation for any yaw configuration may be written as

\begin{equation}
  \sigma_y =
  \begin{cases}
	 \sqrt{y_1^2 + \sigma_{y_1}^2} & \text{if $\gamma_1 = \gamma_2 = \gamma_3 = \gamma_4$ \qquad (equal/zero yaws)}, \\
	 
	\sqrt{y_1^2 + \sigma_{y_1}^2 + \delta_1^2} & \text{if $\gamma_1 = \gamma_2 = -\gamma_3 = -\gamma_4 $ \; (crossed rotors)}, \\
	
	\sqrt{y_1^2 + \sigma_{y_1}^2 + \delta_1^2 + 2\delta_1y_1} & \text{if $\gamma_1 = -\gamma_2 = -\gamma_3 = \gamma_4 $ \; (divergent/convergent rotors)}.
  \end{cases}
\end{equation}
Note that the wake width is not dependent on the sign of deflection in the crossed rotor case, indicating the wake is the similar regardless of which way rotors are crossed. However, this is not the case for divergent and convergent rotors. In these configurations, the wake width is dependent on whether the deflection of the first rotor, $\delta_1$, is positive or negative. If positive, then the wake diverges, whereas if deflection is negative the wake will converge. It is also acknowledged that neither the centroid nor the wake width in the lateral direction is dependent on the vertical spacing or hub height of the turbine, as expected.

\black
\bibliography{article_ref,MR}
\end{document}